\documentstyle[preprint,aps,psfig]{revtex}
\begin{document}

\tightenlines
\title{Cooperativity in Protein Folding: From Lattice Models with Side
Chains to Real Proteins }
\author{D.K.Klimov and D.Thirumalai}

\address{Institute for Physical Science and Technology and 
Department of Chemistry and Biochemistry\\
University of Maryland, College Park, Maryland 20742}

\maketitle

\vspace{1cm}

{\small 

\noindent
{\bf Background: } Over the last few years novel folding mechanisms of
globular proteins have been proposed using minimal lattice and
off-lattice models. However, the factors determining the cooperativity
of folding in these models and especially 
their explicit relation to experiments have not been fully
established. 

\noindent
{\bf Results:} We consider equilibrium folding transitions in lattice
models with and without side chains. A dimensionless measure, \(\Omega
_{c}\), is introduced to quantitatively assess the degree of
cooperativity in lattice models and in real proteins. We show that
larger values of \(\Omega _{c}\) resembling that seen in proteins are
obtained in  lattice models with side chains (LMSC). The enhanced
cooperativity in LMSC is due to the possibility of denser packing of
side chains in the interior of the model polypeptide chain. We also
establish that \(\Omega _{c}\)  correlates extremely well with
\(\sigma = (T_{\theta } -T_{f} )/T_{\theta }\), where \(T_{\theta }\)
and \(T_{f}\) are collapse and folding transition temperatures,
respectively. These theoretical ideas are used to analyze folding
transitions in various two state folders (RNase A, chymotrypsin
inhibitor 2, fibronectin type III modules, and tendamistat) and 
three state folders (apomyoglobin and lysozyme). 
The values of  \(\Omega _{c}\)  extracted from experiments show a
correlation with \(\sigma \) (suitably generalized when
folding is induced by denaturants or acid). For lattice models and for
proteins  \(\Omega _{c} \sim \exp[-(\sigma / \sigma _{0} )^{\beta }]
\), where \( \beta \lesssim 1\). 

\noindent
{\bf Conclusions: } A quantitative description of cooperative
transition of real proteins can be made by lattice models with side
chains. The degree of cooperativity in minimal models and real
proteins can be expressed in terms of 
the single parameter \(\sigma \), which can be estimated from
experimental data. 

\vspace{0.5cm}
\noindent
{\bf Key words: } Lattice models with side chains, cooperativity of folding,
collapse and folding transition temperatures, chemical and thermal
unfolding.  }

\section{\bf Introduction}

In recent years, considerable insight into thermodynamics and  kinetics 
of protein folding has been gained using simple lattice and
off-lattice models \cite{Dill95,Bryn95,Wol,ThirumKlimWood,DillChan97,Mirny}. 
The simplicity of these models allows for
exhaustive sampling of the available conformational space, so that a
detailed picture of the folding mechanisms can be obtained. In almost
all these studies only a very coarse grained description of the
polypeptide chain is retained. These models can be viewed as providing,
in an approximate way, a simple representation of the \(\alpha
\)-carbons of proteins. The amino acid residues  are represented as 
single beads, which is very much in spirit of united atom
approach. Mapping the characteristics of such a coarse grained
description onto real polypeptide chains suggests that each bead may
correspond to 2-3 amino acid residues \cite{Onuchic95,Veit}. 
Lattice models (LM) further restrict
positions of the beads to the vertices of an appropriate
lattice. With the exception of the works by Skolnick and coworkers 
\cite{KGS96} most
of the lattice studies have utilized simple versions of 
cubic and square lattices. 

Despite the success of the minimal models of proteins in providing
general principles of folding mechanisms there are several issues that
have to be resolved before a realistic description of folding can be
provided. In this article, we confine ourselves to one issue, namely,
to what extent is the cooperativity of folding from an unfolded state
to the native conformation mimicked by various minimal models. It is
known experimentally that  under optimal folding
conditions proteins (especially small ones) 
fold in a highly cooperative manner, i.e., they exhibit an 
all-or-none transition \cite{Anfinsen75}. This implies that
there are only two relevant thermodynamic states, namely, the unfolded state
({\bf U}) and the native folded conformation ({\bf N}). The transition {\bf U}
$\rightleftharpoons $ {\bf N} is a first order like phase transition, implying
that there is a barrier (often modest) separating the two
states. The first order nature of folding transition has been seen in
various minimal models of proteins for a variety of interaction schemes. 
Furthermore, it has been shown by various groups
that this transition is a cooperative 
\cite{Dill95,Bryn95,Wol,ThirumKlimWood,DillChan97,Mirny,Cam93,Fukugita93,Brooks97,Karplus97}
depending on a number of factors, 
which are intrinsic to the sequence. However, it has not been
established the degree to which this cooperativity mimics that seen
in real proteins. 

To address this issue quantitatively, we study lattice models with and
without side chains with the purpose of assessing the extent to which
cooperativity of folding transition in real protein  is mimicked by these
models. We introduce a dimensionless measure of cooperativity that
allows us to compare theoretical models and data on real proteins on
nearly equal footing. 
Previous studies have also addressed the factors that
determine the cooperativity of folding using simplified models   
\cite{KGS96,Hao94a,Hao94b}. Hao and Scheraga showed, using a
model similar to that employed here, that cooperativity of folding
transitions results from proper long ranged interactions
\cite{Hao94a,Hao94b}. More
recently, Skolnick and coworkers have used a high coordination lattice
with each side chain having many rotamer degrees of freedom
\cite{KGS96}. They showed that cooperativity, as measured by the
sharpness of the specific heat curves, results from tight side chain
packing \cite{Richards77,Richards93}. 
These results together with our analysis using the    
cooperativity measure clearly show that to
achieve realistic description of thermodynamics one has to include
the rotamer degrees of freedom (at least approximately) 
into coarse grained models so that dense core packing seen in real
proteins can be simulated. We establish that cooperativity can be
linked to experimentally measurable quantities, namely, the collapse
transition temperature \(T_{\theta }\) and the folding transition
temperature \(T_{f}\). Both \(T_{\theta }\) and \(T_{f}\) are
dependent on sequence, pH, and concentration of denaturants.

\section{\bf Methods }

\subsection{\em Lattice Model with Side Chains (LMSC)}

A protein with side chains is modeled on a cubic  
lattice by a backbone sequence of
\(N\) beads, to which a "side" bead, representing a side chain,
is attached. This method of incorporating the rotamer degrees of
freedom in lattice models is similar to that proposed by Bromberg and
Dill \cite{BrDill}. 
In this coarse grained description, the peptide bond and the \(\alpha
-\)carbon are given by a single bead. 
The system has in total \(2N\) beads. The following
types of interactions are included in this model. (i) Self-avoidance,
i.e. any backbone and side beads cannot occupy the same lattice
site more than once. (ii)  The 
interactions between side beads \(i\) and \(j\) \((|j-i| \ge 1)\)  
are taken to be 
pairwise and are assumed to be given by statistical potentials 
derived by Kolinski, Godzik, and Skolnick
(KGS) (ref. \cite{KGS}, Table III). 
We assume that, if two side beads \(i\) and \(j\) \((|j-i| \ge 1)\)  
are nearest neighbors on a lattice, they form a contact and the energy
of their interaction is given by the KGS potential matrix element
associated with the residues \(i\) and \(j\). The interaction energies
in the KGS interaction matrix are in units of \(kT\), where \(T\) is
the absolute temperature. 
In two dimensional representation we have also tried 
Miyazawa-Jernigan (MJ) \cite{MJ85,MJ96} 
parameterization of the matrix of contact 
interactions between   amino acids as well as random site interaction model.

In this study, we took \(N=15\). It is
known that the total number
of symmetry unrelated self-avoiding conformations on a cubic lattice
(i.e., for a model without side chains) is 93,250,730 for  \(N=15\) 
\cite{KlimThirum96}. 
Introduction of side beads  
dramatically increases the  available conformational space, and hence 
exact enumeration of all conformations is no longer feasible even for 
\(N=15\). To illustrate the  sharp increase in the number of available
conformations we take  square lattice as an example. 
On a square lattice the total number of symmetry unrelated
conformations is 296,806 for \(N=15\) \cite{Cam93}. 
When side beads are attached 
the number of available conformations becomes 653,568,850 and the
corresponding number of backbone conformations is 204,988. 
The number of backbone conformations is reduced
because many compact backbone conformations are now inaccessible due
to side chain overlaps. Therefore, inclusion of side chain rotamers
in the model increases the number of conformations by a factor of
2202. In the case of cubic lattice this factor would be even larger
because the coordination number for cubic lattice is 6
instead of 4 for a square lattice. Thus, one has to resort to Monte
Carlo methods to obtain  thermodynamic characteristics of sequences
even for \(N=15\) for cubic lattice models with side chains. The
maximum number of nearest neighbor contacts \(c\) 
in a fully  compact structures
also increases, when side chains are included. In our our previous
work \cite{KlimThirum96} we showed that
for  \(N=15\) \(c\) is  11 in the model without side chains. 
For models with side chains included this
number increases to 22, when only the contacts between side chains are
counted. The maximum total number of nearest neighbor contacts between
backbone beads as well as side chains is 30. 
The larger value of \(c\) gives rise to a
denser packing of the interior, and hence to a greater degree of
cooperativity.

\subsection{\em Selection of Sequences and Determination of Native Structures}

We generated, at random, several sequences using the constraint that the
fraction of hydrophobic residues in a sequence be approximately
0.5 as in the case of naturally occurring proteins. 
Random sequences are
most likely to be not protein-like. In order to obtain viable protein-like
sequences we utilized  the Z score optimization of random sequences
\cite{Eisenberg,Shakh93,Shakh94}  
having unique ground states. The idea of Z score optimization 
is based on minimizing the
relative difference between the energy of a sequence in the  target
conformation \(E_{0}\) and the average energy of misfolded structures
\(E_{ms}\), an estimate for which is \(E_{ms} =
c\!<\!B\!> \), where \(c\) is number of nearest neighbor 
contacts in misfolded structures (presumably equal to the number of
contacts in the native conformation) 
and \( <\!B\!> \) is the average contact energy calculated
for a particular sequence. Thus, the quantity  minimized in
the course of optimization procedure is
\begin{equation} 
Z = \frac{E_{0}-E_{ms}}{\delta}
\end{equation} 
where \(\delta \) is the standard deviation of contact energies for a
given sequence. Standard deviation \(\delta \) and \(E_{ms} \) are
sequence dependent quantities. Low values of Z 
would correspond to sequences with large stability gaps
$(E_{ms}-E_{0})$ \cite{Bryn95}  
and  small \(\delta \).  The sequence with lowest Z score for a
given sequence composition is found by performing Monte Carlo
simulations in  sequence space, the details of which  are
given elsewhere \cite{KlimThirum96,Shakh93,Shakh94}. 
In this study, we used simulated annealing
protocol when doing  Monte Carlo simulations in sequence space and 
we produced a total of 20 trajectories. In the course of simulations we
constantly monitored Z score values and in the end selected sequences
with low values of \(Z\). 

Target conformations used in optimization procedure were randomly
chosen from  preliminary Monte Carlo simulations subject to the
conditions that these structures be reasonably compact, and have a
unique contact map (in order to avoid degeneracy of a ground state). When a
sequence was optimized, we verified that the target conformation 
indeed corresponds to the lowest energy state. To this end, we 
generated 20 slow cooling Monte Carlo
trajectories starting with different initial
conditions at high enough temperature, at which  sequences are in random
coil conformations. The final temperature of these simulations is set
to  0.05, which is well below the folding transition temperature,
\(T_{f}\). We were unable to find any conformation, 
which has lower energy for selected sequence than the
target structure. We also constantly monitored energies in all 
the dynamic simulations. In no instance did we find energies lower than that
corresponding to the native state. The energies and temperature 
in this study are 
expressed in the units of \(kT\) and \(T\), respectively, where \(T\)
is an absolute temperature.

\subsection{\em Probes of Thermodynamics of Sequences: Computation of
\(T_{\theta }\) and \(T_{f} \) }

Thermodynamics of a given sequence was probed using  the overlap
function \(\chi \) \cite{Cam93,KlimThirum96} 
and its fluctuation \(\Delta \chi \) 
suitably modified for the lattice  model with side chains. 
The overlap function, which describes the overall conformation
of a sequence in relation to the native state {\bf N}  is defined as 
\begin{equation}
\chi = 1 - \frac{1}{2N^{2}-3N+1} \Biglb(
\sum_{i < j} \delta(r^{ss}_{ij} - r^{ss, N}_{ij}) + 
\sum_{i < j + 1} \delta(r^{bb}_{ij} - r^{bb, N}_{ij}) + 
\sum_{i \neq j} \delta(r^{bs}_{ij} - r^{bs, N}_{ij}) \Bigrb),
\label{chi}
\end{equation}
where \(r^{bb}_{ij}, r^{ss}_{ij}, r^{bs}_{ij}\) refer  to the
distances between backbone beads, between side beads, and between 
backbone and side
beads, respectively. The factor \(2N^{2}-3N+1\) ensures that in the
native conformation \(\chi = 0\). We also studied the partitioned overlap
functions, \(\chi _{ss}\) and \(\chi_{bb}\), which are defined by
\begin{equation}
\chi_{ss} = 1 - \frac{2}{N^{2}-N} 
\sum_{i < j} \delta(r^{ss}_{ij} - r^{ss, N}_{ij})
\end{equation}
and 
\begin{equation}
\chi_{bb} = 1 - \frac{2}{N^{2}-3N+2} 
\sum_{i < j+1} \delta(r^{bb}_{ij} - r^{bb, N}_{ij}). 
\label{chibb}
\end{equation}
The backbone overlap \(\chi _{bb}\) measures the content of a native
backbone, while the side chain   overlap \(\chi _{ss}\) measures the
"nativeness" of side chain conformations. By treating 
these quantities separately it is easier to monitor 
the onset of native character in the backbone and side chains. 

The acquisition of the overall native state (backbone plus side
chains)  was also probed using the probability
of being in the native structure (a single microstate) 
\begin{equation}
P_{0} = \frac{ \sum_{i} \delta(\chi_{i}) e^{-\frac{E_{i}}{T}} }
{ \sum_{i}e^{-\frac{E_{i}}{T}} }.
\label{P0} 
\end{equation}
In order to observe the acquisition of
the native conformation of the backbone and side beads separately, we 
have calculated the probability that the backbone is in the native 
structure 
\begin{equation}
P_{0}^{b} = \frac{ \sum_{i} \delta(\chi_{i,bb}) 
e^{-\frac{E_{i}}{T}} } { \sum_{i}e^{-\frac{E_{i}}{T}} }
\label{Pb0}
\end{equation}
and, similarly, the  probability of side chains to be in the native
positions  
\begin{equation}
P_{0}^{s} = \frac{ \sum_{i} \delta(\chi_{i,ss} ) e^{-\frac{E_{i}}{T}} }
{ \sum_{i}e^{-\frac{E_{i}}{T}} }. 
\label{Ps0}
\end{equation}
It may seem that there is no need to
consider \(P_{0}^{s}\) separately from \(P_{0}\),  because there are
not many ways of changing 
the backbone conformation while keeping the side chain
contacts intact. Indeed, analysis shows that 
there are very few conformations, which do allow such
changes, but they constitute a small fraction of all structures
so their contribution would be negligible. Our simulations 
show that unless these structures have a considerable Boltzmann weight
(that is highly unlikely) there is no appreciable difference between 
\(P_{0}\) and \(P_{0}^{s}\). 

In Eqs. (\ref{P0},\ref{Pb0},\ref{Ps0}) 
the various probabilities correspond to one
microstate associated with the native conformation, 
which is an artifact of lattice models. This is physically unrealistic
because in any
realistic description there is a volume associated with the native
basin of attraction (NBA). Thus, many conformations will map onto the
NBA. It is, therefore, more relevant to calculate 
the probability of being in the NBA
\begin{equation}
P_{NBA}  = \frac{ \sum_{i} \delta(\chi_{i} \le \chi _{NBA} ) 
e^{-\frac{E_{i}}{T}} } { \sum_{i}e^{-\frac{E_{i}}{T}} }. 
\label{PNBA}
\end{equation}
In Eq. (\ref{PNBA})  the sum is taken over all possible conformations of
backbone and side chains, and \(\chi _{NBA} \)  
is the value of overlap at the temperature, when
the fluctuations of \(\chi \) are maximum, i.e. \(\chi _{NBA} = 
<\chi (T_{f})>\) (see below).   At and below \(T_{f}\) conformations
with \(\chi \le \chi _{NBA}\) will map onto the NBA. It is clear 
that \(P_{NBA}\) 
is physically more relevant than \(P_{0}\). This is because thermal
fluctuations in real proteins frequently 
sample numerous conformations proximal to the native conformation.   

In addition to the functions described above, we calculated the energy
\( <\!E\!> \), the specific heat \(C_{v}\), and the radius of gyration
\(R_{g}\) using standard definitions. 
The collapse transition temperature \(T_{\theta }\) was determined
from the maximum of specific heat \(C_{v}\) \cite{Cam93,KlimThirum96,Veit}
and was found to be well
correlated with the temperature, at which the rate of
change of the radius of 
gyration \(R_{g}\) becomes maximum. As before the folding
transition temperature 
\(T_{f}\) was inferred from the peak of the overlap fluctuation  
\(\Delta \chi \)\cite{Cam93,KlimThirum96,Veit}.

\subsection{\em Calculation of thermodynamic quantities}

The thermodynamic quantities were computed using the multiple
histogram technique \cite{Ferrenberg}, 
which has been implemented in the following way. 
From the slow cooling simulations used  to determine the native conformations 
a rough estimate of collapse and folding transition
temperatures, \(T_{\theta }\) and \(T_{f}\), can be made.   
This allows us to select the temperature range, in which
histograms are to be collected. At each temperature 
histograms were collected using 
\(M=100-200\) Monte Carlo trajectories, each with
different initial conditions. Each
trajectory starts at high temperature \(T_{h} \gtrsim T_{\theta}\) and ends at
the temperature \(T_{l} \simeq T_{f}\). In the
course of a 
trajectory the temperature is changed every \(t\) Monte Carlo steps
(MCS) according to the schedule \(T_{r} =
T_{h} - r \Delta T\), where 
\(r=0,..., (T_{h} - T_{l})/\Delta T \) and the typical value of
\(\Delta T\) is 0.05. Accordingly, histograms were collected
at the temperatures \(T_{r} \) using  \(M\) trajectories
generated. In total, we have \(R=(T_{h} - T_{l})/\Delta T + 1\)
histograms. Five variables were used to collect histograms, these are
overlap function \(\chi \), backbone overlap \(\chi _{bb} \), side
chain overlap \(\chi _{ss} \),
energy \(E\), and the square of the radius of gyration 
\(R_{g}^{2}\).  The length of 
the trajectory at fixed temperature \(T_{r}\) was set constant for all
temperatures and sequences at the value of \(t=2\times10^{7}\) MCS. It is
important to note that a  portion of trajectory immediately
following temperature change must be excluded in order  to allow
the chain to
equilibrate at a new \(T\). We found  this interval of equilibration
does not exceed 
\(2\times 10^{6}\) MCS at the lowest temperature for all sequences. 
For simplicity, the
equilibration interval was kept constant for all temperatures. 
While collecting states in the histograms we used 0.1 grid intervals for
the energy \(E\) and the square of the 
radius of gyration \(R_{g}^{2}\) and exact
discrete values for functions \(\chi \), \(\chi _{bb} \), and 
\(\chi _{ss} \).

The thermodynamic average of a quantity \(A\) that is a   function of,
say, \(\chi \) and \(E\) 
is calculated using multiple histogram method as follows
\begin{equation}
A= \frac{ \sum_{E} \sum _{\chi} A(E,\chi) e^{-\frac{E}{T}}  
\frac{ \sum_{r=1}^{R} h(E,\chi)(T_{r})}
{\sum_{r=1}^{R} n_{r}e^{f_{r}-\frac{E}{T_{r}} } }}
{\sum_{E} \sum _{\chi} e^{-\frac{E}{T}} 
\frac{\sum_{r=1}^{R} h(E,\chi)
(T_{r})}{\sum_{r=1}^{R}n_{n}e^{f_{r}-\frac{E}{T_{r}}}} }
\end{equation}
where \(R\) is the number of histograms, \(T_{r}\) is the
temperature of simulations, at which \(r\)th histogram is collected,
\(n_{r}\) is the number of states in the \(r\)th histogram, and \(f_{r}\)
is the scaled free energy to be calculated self-consistently from
the following equation 
\begin{equation}
e^{-f_{r}} = \sum_{E} \sum _{\chi} e^{-\frac{E}{T_{r}}}
\frac{\sum_{m=1}^{R} h(E,\chi)(T_{m})}{\sum_{m=1}^{R} n_{m}e^{f_{m}-
\frac{E}{T_{m}}}}
\end{equation}
The values of \(f_{r}\) can be determined from an iterative scheme
within  less than 300 iterations with excellent accuracy. 

Using the multiple histogram technique we have calculated overlap function
\( <\!\chi \!> \), its fluctuation 
\(\Delta \chi \), backbone overlap \( <\!\chi_{bb} \!> \), its fluctuation 
\(\Delta \chi _{bb}\), side chain overlap \( <\!\chi_{ss} \!> \), its fluctuation \(\Delta \chi _{ss}\),
energy \( <\!E\!> \), specific heat \(C_{v}\), probabilities
of occupancy of the native state \(P_{0}\), \(P_{0} ^{b}\),
\(P_{0}^{s}\) and the NBA \(P_{NBA}\), 
and the radius of gyration \( <\!R_{g}\!> \) as  functions of temperature. 
From these the key characteristic temperatures \(T_{\theta}\) and
\(T_{f}\) can be readily computed.

\section{Results and Discussion}

\subsection{\em Thermodynamics of sequence A and B}

For the present study we have considered three sequences in three
dimensional (3D) LMSC each with a
unique ground state. In order to keep the description compact we
describe in detail the thermodynamics associated with two sequences, A
and B. The
general characteristics of the third sequence C are similar to those of B. The
compositions of the two sequences along with their native states and
energy spectra are given in Fig. (1).

In Fig. (2) temperature dependencies of various thermodynamic functions 
are plotted for sequence A. The collapse transition temperature
\(T_{\theta }\) is determined from the peak of \(C_{v}\) (see Fig,
(2c)) and 
is equal to 0.27. As shown in Fig. (2c) this temperature coincides with the
temperature, at which the derivative \(d \!<\!R_{g}\!>\!/dT\) shows
maximum. This is a strong evidence that \(T_{\theta }\) indeed 
corresponds to the temperature, at which the 
transition from a random coil state 
to an ensemble of compact structures takes place. 
The plot of \( <\!\chi \!> \) (measuring the content of native state) 
shows that 
equilibrium refolding occurs highly cooperatively, indicating a two
state transition. 
Note that the plot of
\( <\!\chi _{bb}\!> \) follows \( <\!\chi \!> \) very closely. 
It can be seen in Fig. (2b) that the fluctuations in the functions  
\(\chi \) and \(\chi _{bb}\) (as well as \(\chi
_{ss}\),  not shown) 
display a well defined peak at the temperature \(T_{f}=0.26\). 
This value coincides with the temperature, at which 
the probability of being in the NBA is 0.5 (see Fig. (2d)). Thus,
evaluations of \(T_{f}\) from \(\Delta \chi\) and \(P_{NBA}\) are
equivalent.  It is
seen that the backbone fluctuations \(\Delta \chi _{bb}\) 
are stronger at \(T_{f}\)
than the fluctuations of the entire sequence. Thus, for sequence A the
value of parameter \(\sigma _{T}\) is 0.04 (see below). 

It is of interest to compare the probabilities 
\(P_{0}\) and \(P_{0}^{b}\) as a function of temperature. 
We find that the 
native backbone is formed prior to the formation of the entire 
native conformation as the  temperature is lowered (data not shown). In fact, 
the native backbone is predominantly populated at \(T \le T_{0}^{b} =
0.22\), whereas the complete native structure becomes overwhelmingly populated 
only  at \(T \lesssim 0.17\). 
This observation suggests that as the
temperature decreases backbone conformation  becomes frozen, while side chains
still have no well defined conformations seen in the native structure.

Our analysis indicates that sequence A is a two-state folder. 
This is readily inferred from Figs. (2a) and (2b). 
This point
is further illustrated by plotting the thermally weighted 
distribution of (\(\chi, E \)) states at \(T \simeq T_{f}\). 
Such plots can be readily obtained 
from multiple histogram calculations. In
Fig. (3) we plot the two dimensional 
histogram of \((\chi, E) \)   states  for sequence A (upper panel). 
The main feature of this plot
is that it clearly reflects the localization of sequence in two
distinctive states  shown in red  
with very different values of overlaps and energies
- folded (NBA) and unfolded
states. Kinetic simulations (D.K. and D.T., unpublished) 
suggest that the second (unfolded) state
consists of open random coil conformations. Therefore, we classify
sequence A as a two state folder. 

The thermodynamic analysis for sequence B gives \(T_{\theta } =0.44\). 
At roughly the same temperature (=0.48)  \(d\,<\,R_{g}\,>\,/\,dT\)
has a maximum indicating that \(T_{\theta }\) does correspond to the
collapse transition. The value of \(T_{f}=0.30\) and hence \(\sigma =
0.32\). Due to this relatively large \(\sigma \) value we expect the
folding transition to be better described by a multistate
behavior. This is in fact the case and can be seen most clearly from
the two dimensional histogram of \((\chi, E)\) states displayed in the
lower panel of Fig. (3). This plot for sequence B shows that, at
the folding transition temperature, the model polypeptide chain samples at
least three distinctive thermodynamic states shown in red  with 
\(\chi \simeq 0.0\), \(\chi \simeq 0.3\),and \(\chi \simeq 0.65\). 
Notice that the energies at these values of \(\chi \) are also
different. Based on this analysis we expect the folding of B to be
less cooperative than A.

\subsection{\em Comparing cooperativity for LMSC and LM}

The structures of the native states and the corresponding low energy
spectra for the two sequences A and B are given in Fig. (1). Even though
the chain length is small it is seen that  an interior of the
native state of a sequence A is more tightly  packed than that of B.  
By contrast, sequence B has less well formed interior and
hence has a lesser degree of side chain packing. Specifically, the
radius of gyration of a hydrophobic core \(R_{g,H-H}\) (which 
measures the average distance between hydrophobic  residues)
calculated for the native structure of 
sequence A is equal to 1.2. In contrast,
\(R_{g,H-H}\) obtained for sequence B is found to be  1.25. 
It is also interesting to point out that the average distance
between hydrophilic residues  in the native
structure of sequence A is considerably larger (equal to 2.3) 
when compared with that found for
sequence B (equal to 1.9). This suggests 
that the core of sequence A has significant hydrophobic clustering
that leads to denser packing than in sequence B. 
The low energy
spectra of these sequences also reveal different behavior. 
The first excited state of sequence A has the backbone in a non-native
conformation, while the first few excited states of B all have the
native conformation for the backbone. The energy gap (the difference
between the native state and the first excited state) and the so
called stability gap \cite{Bryn95} (the energy difference between the native
conformation and misfolded structures) are larger
for sequence B than for sequence A. Nevertheless, it is shown below
that sequence A folds more cooperatively than sequence B indicating that
the gaps (however they are  defined) are by themselves 
not a good indicator of cooperativity. 

In order to compare the degree of cooperativity in various models and
real proteins we introduce the dimensionless 
cooperativity index defined as 
\begin{equation}
\Omega _{c} = T_{max}^{2} \frac{max[\frac{d\,<\,\chi\,>}{d\,T}]
}{\Delta T},
\label{coop}
\end{equation}
where \(\Delta T\) is the full width at the halfmaximum 
of the peak of \(d\,<\,\chi\,>/d\,T\) 
and $T_{max}$ is the temperature, at which 
\(d\,<\,\chi\,>/d\,T\) has a peak. This quantity can be calculated for
other measures such as the temperature or denaturant dependence of the
fraction of native state, which are experimentally measurable. 

There are other measures of cooperativity that could be used. One such
criterion is based on the van't Hoff equation, from which an index of
cooperativity, namely, \(\Omega _{VH} = \Delta H \Delta T /
RT_{max}^2\) may be introduced. For an infinitely sharp two-state
transition \(\Omega _{VH} \rightarrow const\), while  for a completely
non-cooperative transition  \(\Omega _{VH} \rightarrow 0\). Our
criterion, on the other hand, gives  \(\Omega _{c} \rightarrow
\infty\) for an infinitely sharp two-state transition and tends to
zero, when the transition is non-cooperative. This provides a wider
spread in \(\Omega _{c} \) and we expect it to lead to easier
calibration of cooperativity. Furthermore, when only the dependence of
order parameter (like fraction of native state) as a function of
denaturant concentration is available, a suitable generalization of 
Eq. (\ref{coop}) can be readily
used to compute \(\Omega _{c} \), whereas the generalization of
\(\Omega _{VH} \) to this case is not obvious. 

We have shown elsewhere
\cite{Cam93,KlimThirum96,Veit,KlimThirumPRL} that
folding kinetics (i.e., the time required to reach the native
conformation) is well correlated with 
\begin{equation}
\sigma _{T} = \frac{T_{\theta }- T_{f}}{T_{\theta }},
\label{sigma}
\end{equation}
i.e., sequences with small \(\sigma \) reach the native state more
rapidly than those with larger \(\sigma \). Folding is most rapid near
the tricritical point (\(T_{f} \approx T_{\theta }\)), at which
collapse and acquisition of the native state are almost
synchronous. Parenthetically, we note that  other  
criteria for identifying fast folding sequences have been proposed 
\cite{Bryn95,Mirny}. 

The value of $\Omega _{c}$ for sequence A is
5.32, for sequence B it is 2.03, and for the third 
sequence, C, $\Omega _{c}=1.80$. 
Comparing  $\Omega _{c}$  for sequences A, B, and C shows
that {\em cooperativity is also determined by \(\sigma \)};  \(\sigma
\) for A is 0.04, for B is 0.32, and for C is 0.34. This is shown (see below) 
to be the case for both LMSC and LM. 

It is interesting to compare  $\Omega _{c}$ for lattice
models with side chains and lattice models   without
side chains. For this, we consider a number of 15-mer sequences without
side chains on a cubic lattice. The sequences used for this purpose 
are the ones 
studied in our previous papers \cite{KlimThirum96,KlimThirumPRL}. 
Many of these sequences fold cooperatively
with \(\sigma \) values ranging from 0.0 to 0.79. In Fig. (4) we plot 
$\Omega _{c}$ as a function of \(\sigma \) for 15-mer lattice
sequences. The maximum value of $\Omega _{c}$ of 3.0 is observed for
sequences with \(\sigma \simeq 0.0\). This level of cooperativity is
less than that seen in the models with side chains indicating that
cooperativity in natural proteins results from the dense packing of
side chains in the core \cite{KGS96,Richards77}. Fig. (4)  
shows remarkable statistical  
correlation between the cooperativity index $\Omega _{c}$
and \(\sigma \). When combined with our previous results we can
conclude that both cooperativity and folding kinetics is determined by
the intrinsic thermodynamic parameters \(T_{\theta }\) and \(T_{f}\). 
It should be emphasized that \(T_{\theta }\) and \(T_{f}\) are
sensitive to mutations as well as external conditions such as pH,
ionic strength, etc. 

In order to further establish the correlation between 
$\Omega _{c}$ and \(\sigma \) we have studied two
dimensional (2D) LMSC using a number of sequences with very different
interaction potentials (see caption to Fig. (5)). 
Sequence properties have been inferred by
exhaustive enumeration of all allowed conformations. This permits us
to analyze much  larger sequence database than for 3D 
LMSC. In Fig. (5) we plot $\Omega _{c}$  s a function of \(\sigma \)
for our 2D models with side chains. As is the case for \(N=15\) 3D LM the
correlation is excellent. It should be noted that $\Omega _{c}$ values
are considerably smaller than in 3D indicating that for realistic
depiction of cooperativity 
2D simulations may not be reliable.

\subsection{\em Dependence of $\Omega _{c}$ on chain length }

The model with side chains for \(N=15\) yields $\Omega _{c}$
values of the order of 5, when \(\sigma \) is small, whereas real
proteins, which are  two state folders, have $\Omega _{c}$ in excess of
10 (see below). This discrepancy between LMSC  and experiments is due
probably to
the small value of \(N\). If \(N\) is increased, then we expect to
obtain $\Omega _{c}$ in the experimentally measured range. This
assertion is borne out by examining $\Omega _{c}$ for lattice models
without side chains for larger \(N\). In Fig. (6) we display $\Omega
_{c}$ vs \(\sigma \) for \(27\)-mer sequences, for which detailed
account of thermodynamics and kinetics was provided in a previous study
\cite{KlimThirum96}. For \(N=27\) we
see that $\Omega _{c}$ ranges from 1.0 (\(\sigma = 0.35\)) to 33.3
(\(\sigma =0.0\)). This figure also shows that cooperativity is
essentially determined by \(\sigma \). The results in Figs. (4-6) show
that one can write 
\begin{equation}
\Omega _{c} \sim \exp [-(\sigma/\sigma _{0})^{\beta }], 
\end{equation}
where \(\sigma _{0}\) is a model dependent constant and  
\(\beta \lesssim 1\). 
Since the addition of side
chains enhance $\Omega _{c}$, we conclude that in order to
provide a realistic description of cooperativity in protein folding it
is necessary to include side chains in minimal models of proteins.

\subsection{\em Extracting $\Omega _{c}$ and \(\sigma \) from experiments }

We have computed  $\Omega _{c}$ for a few
proteins and a protein fragment using experimental 
data such as the dependence of
fraction of native state as a function of temperature, denaturant or
acid \cite{CI2,RNase,tend,FNIII,barnase,apoMb,lysozyme}. 
The proteins we chose for our analysis are for
illustrative purposes only and does not constitute an
exhaustive analysis. 
The data from the literature when used in combination with
Eq. (\ref{coop}) gives $\Omega _{c}$  in the range 0.25 - 99.4 (see
Tables I-III). Out of the eight 
proteins considered here the fragment of barnase
is least cooperative \cite{barnase}. 
This, in fact, suggests that the cooperativity is
greatly compromised in the absence of long range interactions with the
rest of the protein. Surprisingly, the cooperativity of the ninth
module of fibronectin exhibits the level of cooperativity similar to
the barnase fragment. 
Chymotrypsin inhibitor 2 (CI2), RNase A, tendamistat,  
and the tenth module of fibronectin ($^{10}$FNIII) have been recently
shown \cite{CI2,RNase,tend,FNIII} 
to fold in a very cooperative two state manner. 
Apomyoglobin (apoMb) and lysozyme \cite{apoMb,lysozyme} 
have been shown to be three state
folders. Below we describe the cooperativity of these proteins using
the equilibrium folding data. 

\noindent
{\bf Two state folders: } The thermal unfolding of RNase A and RNase B
has been recently reported \cite{RNase}. The data for the fraction of native
state as a function of temperature (see Fig. (5) of \cite{RNase}) clearly
exhibits a cooperative two state transition. 
Using the parameters
obtained from the two state analysis of  RNase A in \cite{RNase} and 
Eq. (\ref{coop}) (where \(<\chi >\) is replaced by the native fraction
\(f_{N}\)) we get $\Omega _{c} \approx 99.4$. The large value of 
$\Omega _{c}$  makes us suggest that for this protein \(\sigma \approx
0.0\) implying that \(T_{f} \approx T_{\theta }\). A more precise 
estimate  of \(\sigma \) is given below. 

Kiefhaber and coworkers \cite{tend} 
have argued using the ellipticity measurements
at pH=7 and \(T=25^{\circ}C\) that tendamistat (a small protein
consisting of 74 amino acid residues) unfolds thermodynamically in
an all-or-none fashion upon increasing the concentration of guanidinium
chloride. The value of $\Omega _{c}$ calculated using 
the dependence of the fraction of native state \(f_{N}\) (see Eq.(\ref{fn})) 
on the concentration of GdnHCl is 13.7, which is considerably
less than for RNase A. This suggests that either \(\sigma \) for
tendamistat is larger than for RNase A or that  $\Omega _{c}$ is
intrinsically smaller when folding is induced by varying the
concentration of denaturants as opposed to thermal unfolding. The
determination of \(\sigma \) (see below)  suggests that the latter is a more
likely possibility. 

Recently, Dobson and coworkers have shown that 
denaturation induced unfolding of fibronectin type
III $^{9}$FNIII and $^{10}$FNIII 
proteins is well described by a two state model \cite{FNIII}. 
However, it was found that 
$^{9}$FNIII has a  lower stability (as measured by a free energy
difference between {\bf U} and {\bf N}) than $^{10}$FNIII. 
The free energy of unfolding in water \(\Delta G _{H_{2}O}
= 1.2 \) and 6.1 \(kcal\,mol^{-1}\) for $^{9}$FNIII and $^{10}$FNIII, 
respectively. In line with these observations, 
our analysis suggests that both proteins display sharply different
degrees of cooperativity with \(\Omega _{c} = 0.28\) obtained for   $^{9}$FNIII
and 7.5 for  $^{10}$FNIII. Accordingly, we expect that 
the \(\sigma \) values are very different for $^{9}$FNIII and
$^{10}$FNIII.  

A few years ago Jackson and First \cite{CI2} reported that folding of CI2
proceeds via a two state manner. This was the case whether
denaturation was induced chemically or thermally. By using 
fluorescence, \(F\), as a function of [GdnHCl] they 
extracted  parameters for the two-state description of the data. 
Since the values of \(F_{N}\) and \(F_{U}\)
(fluorescence in the native state and denaturated state, respectively)
were not reported we determine the fraction of native state for
chemically induced unfolding using 
\begin{equation}
f_{N} = \frac{F-F_{N}}{F_{U}-F_{N}} = 
\frac{1}{1+\exp^{ -\frac{\Delta G_{U}}{RT} } },
\label{fn}
\end{equation}
where \(\Delta G_{U} = \Delta G_{H_{2}O} - m[GdnHCl]\) and \(\Delta
G_{H_{2}O}\) is the free energy of unfolding in water, \(m\) is  a
constant that effectively measures the degree of exposure of the
protein upon denaturation, and \(R\) is the gas constant. 
For Eq. (\ref{fn}), which gives a two-state description of denaturant
induced unfolding, \(\Omega _{c}\) can be shown to be
\begin{equation}
\Omega _{c} = \frac{1}{4}\Biggl( \frac{\Delta G_{H_{2}O}}{RT} \Biggr)^{2}
\frac{1}{ln \frac{3+2\sqrt{2}}{ 3-2\sqrt{2} } },
\label{Omega_theor}
\end{equation}
implying that cooperativity in this case is directly related to
stability \cite{Baldwin97}.  
Using the experimental parameters \(\Delta
G_{H_{2}O}=7.03\, kcal\, mol^{-1}\) and 
\(m=1.79\, kcal\, mol^{-1}\, M^{-1}\) \cite{CI2} we
find $\Omega _{c} = 9.9$ - a value that is comparable to tendamistat.

Jackson and Fersht also reported 
thermally induced unfolding of CI2 as
a function of pH \cite{CI2}. They measured the pH dependence of the folding
transition temperature \(T_{f}\) (referred to as \(T_{m}\) in
ref. \cite{CI2}) using scanning microcalorimetry experiments. They showed
that in the range of pH (2.2\,-\,3.5) CI2 unfolds thermally also by a two state
manner. We calculated $\Omega _{c}$  using the data given in \cite{CI2}
and find that it lies between 5.9 at pH=2.2 to 29.3 at pH=3.5. The
degree of cooperativity clearly increases with pH. At the higher
values of pH $\Omega _{c}$ is larger than that found by chemical
denaturation. Thus, in general, it appears that the degree of
cooperativity is greater when unfolding is induced thermally than by
adding denaturants.

We have established in the simulations on lattice models that $\Omega
_{c}$ is well correlated with \(\sigma _{T}\) (see Figs. (4-6)). 
In order to verify this in experiments one needs measurements of
\(T_{\theta }\), which are not currently available. So we have devised
a way to obtain \(T_{\theta }\) using the following arguments. 
Since
there are two separate phase transitions at \(T_{\theta }\) and
\(T_{f}\) the experimental
determination of \(\sigma \) requires separate  measurements of
order parameters, probing these "phases".  The estimate of 
\(T_{\theta }\) requires independent measurement of a quantity
associated with the compaction of a polypeptide chain (for example,
the radius of gyration, \(R_{g}\)). At present, simultaneous 
measurements of \(R_{g}\) and the fraction of native state \(f_{N}\)
are extremely rare \cite{lysozyme}. In the absence of such measurements we have
used an ad hoc empirical procedure  to
estimate \(\sigma \) from the existing data so that the variation of
$\Omega _{c}$ with pH, such as that seen in CI2, can be rationalized. To
estimate \(T_{\theta }\) from experiments we propose the following
approach. We calculated \(T_{\theta }^{\prime}\) (an estimate for
\(T_{\theta }\))  using the condition 
\begin{equation}
f_{N}(T_{\theta }^{\prime}) = C,
\label{tc}
\end{equation}
where \(C\) here is taken to be 0.05. 
At \(T_{\theta }\) the equilibrium is shifted almost to the unfolded
state and, consequently, we expect \(C\) in Eq. (\ref{tc}) to be small. 
Recall that \(T_{f}\) is
determined in the analysis  of experimental data from
\(f_{N}(T_{f}) = 0.5\). 

In order to justify the use of Eq. (\ref{tc})
we have estimated \(T_{\theta }^{\prime}\) using 
\(P_{0}(T_{\theta }^{\prime})=0.05\)
for LM with \(N=27\).  A plot of \(T_{\theta }^{\prime}\) from this estimate
as a function of actual \(T_{\theta }\) (from the peak of the specific
heat) is given in Fig. (7). There is a very good correlation (with the
slope being near unity) with the
quality getting better as \(T_{\theta }\) becomes larger, which
corresponds to small \(\sigma \) values. A similar correlation is found
when \(C\) is changed to 0.1. This method of estimating \(T_{\theta
}\), although arbitrary, is physically motivated. We find
that if \(C\) is increased beyond 0.1 the correlation coefficient
considerably decreases. 
Thus, this way of obtaining  
\(T_{\theta }\) from experimental data  (with \(C \lesssim 0.1 \))
is useful especially for the two state folders. 

Additional justification of the use of Eq. (\ref{tc}) to
estimate \(T_{\theta }\) can be given by examining the temperature
dependence of \(R_{g}\) for RNase A. Sosnick and Trewhella have used
small X-ray scattering to measure \(R_{g}\) as a function of
temperature (see Fig. (2) of \cite{Sosnick92}). For the
nonreducing conditions, which were the ones considered in \cite{RNase}
as well, 
we estimate \(T_{\theta }\) (from Fig. (2) in  \cite{Sosnick92}) to be
around 66$^{\circ }$C. This value is remarkably close to that obtained
using Eq. (\ref{tc}) (\(T_{\theta } \simeq 65.2^{\circ} \)C) 
using the data of Ulbrich-Hofmann and
coworkers. We should emphasize that despite the reasonable  estimates
of \(T_{\theta }\) from  Eq. (\ref{tc}) independent experimental
measurements are needed. 

We have used Eq. (\ref{tc}) together with the data on thermal
unfolding of CI2 and RNase A to estimate 
\(T_{\theta }\). These are given in Table I along with their $\Omega _{c}$
and \(\sigma \). 
It is seen from Table I that $\sigma _{T} $ is relatively large 
for CI2 at pH=2.2. From the previously
established results connecting folding kinetics and \(\sigma \)
\cite{KlimThirum96,KlimThirumPRL,Veit} we
predict that there should be a significant off-pathway process in the
refolding of CI2 at pH=2.2 at \(T \approx T_{f}\). It would be
interesting to determine the pH dependence  (over a wider range) of
\(T_{f}\) for the variant of CI2 used in the recent experiments of
Fersht and coworkers, which contains 64 amino acid residues as opposed
to 81 in the original work of Jackson and Fersht. 

In most of the experiments the equilibrium unfolding is achieved by
chemical denaturation. The analysis of such experiments can be done by
generalizing \(\sigma \). We introduce 
\begin{equation}
\sigma _{D} = \frac{C_{\theta } -C_{f}}{C_{\theta } },
\label{newsigma}
\end{equation}
where $C_{\theta }$ is the concentration of denaturant, at which the
transition to compact structures from the random coil state takes
place, and \(C_{f}\) is the denaturant concentration in the midpoint of
an appropriate measure of the native state population. Similarly,
\(\Omega _{c}\) in Eq. (\ref{coop}) can be 
calculated using the denaturant dependence of the fraction of native
state. With this generalization we have estimated \(C_{f}\)
and $C_{\theta }$ for tendamistat, $^{9}$FNIII, $^{10}$FNIII, and 
CI2 (denaturant induced unfolding). 
The value of $C_{\theta }$ is obtained from the condition
\(f_{N}(C_{\theta }) =0.05\). The values of \(\sigma \) and \(\Omega
_{c}\) are given in  Table II. 

The physical basis for proposing Eq. (\ref{newsigma}) as a natural
generalization of Eq. (\ref{sigma}) is the following. When the
temperature changes from below \(T_{f}\) to above \(T_{\theta}\), 
there are two phase changes, one at \(T_{f}\) and the other at
\(T_{\theta }\). Similar phase changes are seen upon chemical
denaturation. Both \(\sigma _{D}\) and \(\sigma _{T}\) capture the
physics that fastest folding should occur when the intermediate is
eliminated. Quantitatively, the value of \(\sigma _{D}\) and \(\sigma
_{T}\) need not coincide because the nature of interactions due to
thermal unfolding and denaturant induced unfolding is different. In
fact, it appears from the analysis of the experiments in this study
that thermal denaturation is sharper than chemically induced
unfolding.

\noindent
{\bf Three State Folders: } Barrick and Baldwin \cite{apoMb} showed that
urea and acid induced folding of apomyoglobin (apoMb) is best described by
a three state model. Thus, the equilibrium folding is given by 
\begin{equation}
{\bf U } \rightleftharpoons {\bf I } \rightleftharpoons {\bf N }. 
\end{equation}
Here we analyze the acid induced transition using the data presented
in Fig. (1) and Table I of \cite{apoMb}. In Fig. (1) of \cite{apoMb}  Barrick
and Baldwin present the helical content of apoMb as a function of pH
at various urea concentrations. We have ignored the data at 4.5 M
concentration of urea, because there appears to be no significant
change in \([\theta ]_{222}\) with pH  as measured   by CD spectroscopy. At 3.0
M the value of  \([\theta ]_{222}\) at the highest pH is considerably
less than at all other values of urea concentration, which does not seem to be
reflected in the quantitative analysis presented in Table I of
\cite{apoMb}. Examination of all other curves clearly shows evidence for
three state behavior. If we calculate \(\Omega _{c}\) for the  transition
${\bf U } \rightleftharpoons {\bf N }  $, ignoring the equilibrium
intermediate {\bf I}, we find that the overall cooperativity is
highly compromised. For the purposes of measuring the cooperativity of
folding we have focused separately on the transitions 
${\bf U } \rightleftharpoons {\bf I }  $ and 
${\bf I } \rightleftharpoons {\bf N }  $ and calculated 
\(\Omega _{c} ({\bf U } \rightleftharpoons {\bf I })\) and 
\(\Omega _{c} ({\bf I } \rightleftharpoons {\bf N })\)
using a simple generalization of Eq. (\ref{sigma}). An estimate of
\(\sigma _{pH}\) can be easily made by associating the midpoint of the
transition ${\bf U } \rightleftharpoons {\bf I }  $ with the collapse
process and the midpoint of 
the transition ${\bf I } \rightleftharpoons {\bf N }  $
with the folding transition. The values of \(\Omega _{c}\) and
\(\sigma _{pH}\)  for acid induced refolding are given in Table
III. It is clear that \(\Omega _{c}\) is largest for the smallest
values of \(\sigma _{pH}\), which is consistent with the results for
LMSC and LM and the two state folders. In Table III we also present
the cooperativity index \(\Omega _{c} ({\bf U } \rightleftharpoons
{\bf I })\) for apoMb. It is clear from Table III that 
\(\Omega _{c} ({\bf U } \rightleftharpoons
{\bf I })\) is considerably less than 
\(\Omega _{c} ({\bf I } \rightleftharpoons {\bf N })\) at all values
of urea concentration. The very small values of \(\Omega _{c} ({\bf U } \rightleftharpoons
{\bf I })\)  suggest that the initial compaction
results in the formation of an equilibrium intermediate, whose structural
characteristics are closer to {\bf U} than to {\bf N}.

Further evidence for our assertion that 
${\bf U } \rightleftharpoons {\bf I }  $ transition in apoMb is
less cooperative (as measured by \(\Omega _{c}\)) comes from the recent
work of Luo {\em et al. } \cite{Baldwin97}. These authors measured the
unfolding thermodynamics from the pH 4 folding intermediate of
apoMb, in which A, G, and H helices are ordered. By using mutations in
the A and G helices Luo {\em et al. } argue that the unfolding from
{\bf I} is well described by a two state thermodynamics. Although the
authors did not explicitly provide \(\Delta  G_{H_{2}O}\) values
(needed in Eq. (\ref{fn})) we scanned the data in Fig. (3) of \cite{Baldwin97}
and calculated \(\Omega _{c}\) values. The resulting \(\Omega _{c}\)
values for ${\bf U } \rightleftharpoons {\bf I }  $ transition is
given in Table III. We find that these values are remarkably
consistent with those obtained from the pH induced unfolding
transition ${\bf U } \rightleftharpoons {\bf I }  $ reported in 
\cite{apoMb}. Thus, it appears that the assembly
of the native state from the equilibrium intermediate {\bf I} is more
cooperative than the formation of {\bf I} itself from the unfolded state. 

Doniach and coworkers have argued using radius of gyration data that
folding of lysozyme at pH=2.9 is better described by a three state
analysis rather than by an all-or-none transition \cite{lysozyme}. 
We have used the theoretical methods developed here to
compute \(\sigma _{D}\) and \(\Omega _{c}\). These are displayed in Table
III. The value of \(\sigma _{D}\) is smaller than for apoMb. Unlike the
results for apoMb the data for denaturant induced unfolding of
lysozyme does not exhibit a clear evidence of a third equilibrium
intermediate at pH=2.9. This observation together with relativity
small value of \(\sigma \) makes us suggest that at pH=2.9 lysozyme is,
at best, a borderline  three state folder. 

In Fig. (8) we plot  \(\Omega _{c}\) as a function of \(\sigma \)
using the data given in Tables I, II, and III. This figure shows a
good correlation between two quantities. 
If the transition is strictly 
an all-or-none, then from a
mathematical representation of the resulting sigmoid curve one can
show that the correlation between \(\Omega _{c}\) and \(\sigma \)
should follow independent of the value of \(C\) used in Eq. (\ref{tc})
to get \(T_{\theta }\) (and hence \(\sigma \)). Thus, the good
correlation between \(\Omega _{c}\) and \(\sigma \) for two state
folders is not surprising. Fig. (8) also shows a good correlation for 
acid induced unfolding of apoMb, 
which is a three state folder (see Table III). For this protein, 
\(T_{\theta }\) and \(T_{f}\) are calculated
from the midpoints of ${\bf U } \rightleftharpoons {\bf I }  $ and
${\bf I } \rightleftharpoons {\bf N }  $ transitions, respectively. 
Thus, we feel that the degree of
cooperativity in protein folding can be understood in terms of 
\(\sigma \) with \(T_{\theta }\) being empirically defined by
Eq. (\ref{tc}). The physically motivated choice for \(C\) (\(\lesssim
0.1\)) is further supported by the relationship between folding times
and \(\sigma \) for several proteins (D.K. and D.T., unpublished). 

It can be seen from Table II that there is a good correlation between
\(\Omega _{c}\) and \(\Delta G _{H_{2}O}\), which is the free energy
difference between the {\em unfolded state } and the native state. The
value of \(\Delta G _{H_{2}O}\) is normally obtained by extrapolating
to zero denaturant concentration which involves errors. More
importantly, this analysis of correlation between  \(\Omega _{c}\) and 
\(\Delta G _{H_{2}O}\) cannot be carried out easily in the case of  thermal
unfolding. As a result it appears that it is more
convenient to analyze cooperativity data in terms of \(\sigma \), which
can be estimated using our empirical criterion 
for both thermal and chemical unfolding transitions.

By comparing the range of \(\sigma \) values for two and three state
folders (see Tables I, II, III) we can conclude that proteins with 
\(\sigma \lesssim 0.25\) should exhibit an all-or-none transition. If 
\(\sigma \gtrsim 0.25\), then one should anticipate a three (or more)
state folding transition. The value of \(\sigma \sim 0.25\), at which
crossover from two state behavior to a three state transition takes
place, should be viewed as  a rough estimate.

\section{Conclusion}

One of the most important characteristics of protein folding is that
the equilibrium folding transition is cooperative \cite{Anfinsen75}. 
The minimal models (lattice and off-lattice)
without side chains capture this cooperativity only in a qualitative
manner \cite{Dill95,Bryn95,Wol,ThirumKlimWood,DillChan97,Mirny}. 
If a quantitative description of this process is required, then,
as demonstrated here and elsewhere \cite{KGS96}, 
a coarse grained description
of side chains becomes necessary. Inclusion of side chains into a lattice
model of even a small sized chains gives enhanced cooperativity. In
these more realistic models cooperativity is due to  the
possibility of denser packing \cite{Richards77} of the interior of the folded protein. It is
clear from the results for \(N=27\) without side chains that
the degree of cooperativity 
approaching that seen in two state folding of real proteins can be
easily achieved by including side chains. 

We have shown that there is a correlation between the
measure of cooperativity $\Omega _{c}$ (see Eq. (\ref{coop})) and
\(\sigma  = (T_{\theta} - T_{f})/T_{\theta} \). 
Sequences with small \(\sigma \) fold in a two state manner with large
values of $\Omega _{c}$. This is the case for lattice models as well as
real proteins. 
The estimates of \(\sigma \) for two state folders were made using the
parameterization of experimental data, measuring the fraction of the
native state. Since there are two transitions involved it would be
desirable to obtain \(T_{\theta }\) using entirely different probe
(for example, by measuring the radius of gyration). Such measurements
would be required to firmly establish the relationship between
\(\Omega _{c}\) and  \(\sigma \). In the absence of such measurements
our estimate of \(\sigma \) for two state folders based on
Eq. (\ref{tc})  is expected to be only approximate. 

Proteins, which require a three state model to describe
their folding transition, are expected to have relatively large
\(\sigma \). Based on this classification, we expect  that for these
proteins the actual cooperative transition should occur from an
equilibrium intermediate to the native state as the temperature or
denaturant concentration is altered. Such a cooperative transition 
has in fact been
recently reported in the folding of apobovine \(\alpha -\)lactalbumin 
\cite{Balbach}. The
small values of $\Omega _{c}$ 
for the overall ${\bf U } \rightleftharpoons {\bf N }  $ transition 
(presumably, resulting from large values
of \(\sigma \)) for lysozyme and apoMb is a reflection of a poor
cooperativity of the transition from  a random coil to the native
state due to the presence of the equilibrium intermediates.  
This is consistent with a small values of \(\Omega _{c}\) for
transitions ${\bf U } \rightleftharpoons {\bf I }  $, while
transitions   
${\bf I } \rightleftharpoons {\bf N }  $ are highly cooperative. 

The detailed study of the transition to the native state from unfolded
conformations in lattice models with side chains suggests a clear
mechanism for folding of small proteins that exhibit   all-or-none
transitions. At
a relatively high temperature the backbone becomes "frozen" adopting a
native conformation, but side chains are still mobile. Only at
lower temperature the side chains become predominantly fixed. The
temperature interval, over which this happens, is small for sequences
with small \(\sigma \), so that these sequences are expected to fold
kinetically and thermodynamically in a two state manner. 
This implies that for these sequences pathways should exist, in which
the acquisition of the native conformation (backbone and side chains)
and the collapse  process should be almost simultaneous. Recent
molecular dynamics simulations on small peptides also support this
notion \cite{Mohanty}. It is clear that 
the enhanced degree of cooperativity in LMSC
over LM is due to packing of side chains. A corollary of this result is that
denaturation in proteins is basically caused by the disruption of the
densely packed side chains as was suggested some years ago by
Shakhnovich and Finkelstein \cite{Finkel}.

Sequences with moderate or large \(\sigma \) have large difference
between 
the temperatures at which the backbone and side chains adopt
native-like conformations. As a result these sequences get trapped in
an intermediate, which inevitably  slows down the folding kinetics. 
Explicit
kinetic simulations of models with side chains confirm these 
assertions (D.K. and D.T., unpublished). 

\acknowledgments

We would like to thank Prof. R. Ulbrich-Hofmann for providing the thermal
unfolding data for RNase A in tabular form. We are grateful to
W.A.Eaton for suggesting Eq. (\ref{newsigma}) as a generalization of
Eq. (\ref{sigma}). We thank J.D.Bryngelson for pointing out
reference \cite{Sosnick92} and T.R. Sosnick for useful discussions. We are also
grateful to the referees for penetrating comments.  
This work was supported in part by  a grant from 
the National Science
Foundation (through grant number  CHE96-29845).

\begin{figure}

\noindent
{\bf Fig. 1.} Native conformations and  energy spectra 
of sequences A (upper panel) and B (lower panel). 
The sequences for A and B are also displayed. The first letter
corresponds to residue labeled 1. 
The energy spectra for the two sequences is given on the right. The
spectra for the two sequences are divided into two columns. The left
column gives the low energy levels for conformations, in which the
backbone (and hence most likely the entire structure) is in a
non-native state. The right column corresponds to conformations with
the backbone in the native state. 
Sequence A is highly optimized with 
\(Z=-17.1\) and \(\delta = 0.50\), whereas sequence B, 
while having approximately the same Z score factor of
\(-17.9\), is not  well optimized. This is because one
can  design a sequence with the {\em same } composition as sequence B
has but with smaller \(Z\). Consequently, the value of \(\delta =0.74\) for
sequence B is larger, which indicates a higher degree of 
heterogeneity of this sequence. 
Sequence A has 10 hydrophobic residues, while B has 8. The figure has
been generated using RasMol program (R. Sayle, 1995).

\noindent
{\bf Fig. 2.} Various thermodynamic functions as a function of
temperature for sequence A. Fig. (2a) shows \(<\chi >\) (solid line)
and  \(<\chi _{bb}>\) (dashed line) (see Eqs. (\ref{chi}) and (\ref{chibb}),
respectively). The fluctuations   \(\Delta \chi \)  (solid line) and 
 \(\Delta \chi  _{bb}\)  (dashed line) are shown in Fig. (2b). The
temperature dependence of the specific heat \(C_{v}\) (solid line) and
the derivative of the radius of gyration with respect to temperature 
\(d<R_{g}>/dT\) (dashed line)  are displayed in
Fig. (2c). The value of temperature corresponding to the peak in
\(C_{v}\) is associated with \(T_{\theta }\), which clearly coincides
with the peak in \(d<R_{g}>/dT\). This demonstrates that \(T_{\theta
}\) is the usual collapse transition temperature. 
The probability of
being in the NBA \(P_{NBA}\) as a function of temperature is given in
Fig. (2d) with the horizontal line marking the level of 0.5.  The
temperature, at which \(P_{NBA}\) is 0.5, is 0.25, which is almost the
same value, at which \(\Delta \chi \) has a maximum. Thus, two methods
for computing \(T_{f}\) are equivalent.

\noindent
{\bf Fig. 3.} Thermally weighted distribution of states \((\chi, E) \) 
calculated for sequence A (upper panel) 
and sequence B (lower panel). The most populated states \((\chi, E) \)
are shown in red. This figure clearly shows that, at
the folding transition temperature \(T_{f}\),  
sequence A samples two distinct thermodynamic
states - native basin of attraction and  the random coil (unfolded)
states. Sequence B, on the other hand, samples at \(T_{f}\) at least
three distinct states with different \(\chi \) and \(E\). We expect the
thermodynamic transition for sequence A to be two state like, whereas
sequence B is better described by a three state analysis.

\noindent 
{\bf Fig. 4} Dependence of cooperativity \(\Omega _{c}\) on 
\(\sigma _{T} \) for 
3D \(15-\)mer sequences without  side chains (full circles) and 
with side chains (empty circles). Solid line is a guide to
eye. 

\noindent 
{\bf Fig. 5} Dependence of cooperativity index \(\Omega _{c}\) on 
\(\sigma _{T} \) for 2D \(15-\)mer sequences with side chains and different
interaction matrices: random site model (diamonds), MJ
shifted potentials (full circles), MJ potentials (empty circles), KGS
shifted potentials (triangles), KGS
potentials (squares). Shifted MJ or KGS potentials are obtained by
setting \(<B_{0}>=0\), where \(<B_{0}>\) is the interaction energy
averaged over all pairs of residues. 
The values of \(\sigma _{T}\) and \(\Omega _{c}\) 
are obtained by exact enumeration. Solid line is 
a guide to eye.

\noindent 
{\bf Fig. 6} Dependence of cooperativity index \(\Omega _{c}\) on 
\(\sigma _{T}\) for 3D \(27-\)mer sequences without  side chains. 
The solid line is a guide to  eye. 

\noindent 
{\bf Fig. 7} Dependence of the collapse temperature \(T_{\theta }^{\prime}\)
estimated using Eq. (\ref{tc}) on actual collapse temperature
\(T_{\theta }\) determined from the peak of specific heat \(C_{v}\)
for 27-mer LM sequences. The solid line is
a guide to  eye. 

\noindent 
{\bf Fig. 8} Variation  of the cooperativity 
index \(\Omega _{c}\) on \(\sigma \) for CI2 (thermal unfolding - full
circles, chemical unfolding - empty circle), tendamistat (square), and
RNase A (triangle), $^{10}$FNIII (star), lysozyme (diamond), and apoMb
(crosses). 
We did not show the results for the barnase fragment and $^{9}$FNIII, 
because  \(\Omega _{c}\) is too small for them. 
This figure shows that cooperativity can be parameterized by 
\(\sigma \), which can be
experimentally determined.

\end{figure}

\newpage
\begin{table}
\caption{Temperature Induced Unfolding: Two State Folders }

\begin{tabular}{lccccc}
     & \multicolumn{2}{c}{experimental values} &  
\multicolumn{3}{c}{theoretical estimates}\\ \hline 
Protein & pH & \(T_{f}, ^{\circ}C \) & \(T_{\theta }, ^{\circ}C ^{c}\) & \(\sigma \) & \(\Omega _{c}\) \\ \hline

CI2$^{a}$       & 2.2 &   41.5  &  55.5  &    0.25    &   5.9   \\ 
                & 2.5 &   47.1  &  59.8  &    0.21    &   9.1   \\ 
                & 2.8 &   55.1  &  67.5  &    0.18    &  13.1   \\ 
                & 3.2 &   63.9  &  75.2  &    0.15    &  21.2   \\ 
                & 3.5 &   70.9  &  81.5  &    0.13    &  29.3   \\ \hline
RNase A$^{b}$   & 8.0 &   60.3  &  65.2  &    0.08    &  99.4   \\ \hline

\end{tabular}
\end{table}

\noindent
$^{a}$ reported in \cite{CI2}\\
$^{b}$ reported in \cite{RNase}\\
$^{c}$ estimated using Eq. (\ref{tc})\\

\newpage

\begin{table}
\caption{Chemically Induced Unfolding: Two State Folders }

\begin{tabular}{lccc}
 & experimental values &  \multicolumn{2}{c}{theoretical
estimates}\\ \hline 

Protein  &  \(\Delta G _{H_{2}O}, kcal\,mol^{-1}\) & \(\sigma \) & \(\Omega _{c}\) \\ \hline

Tendamistat$^{a}$       &  8.13   & 0.17 &  13.7 \\ 
CI2$^{b}$               &  7.03   & 0.20 &  9.9 \\ 
$^{10}$FNIII $^{c}$     &  6.1    & 0.22 &  7.5  \\
$^{9}$FNIII $^{c}$      &  1.2    & 0.60 &  0.28  \\
Barnase fragment$^{d}$  &  1.1    & 0.62 &  0.25 \\  \hline

\end{tabular}
\end{table}

\noindent
$^{a}$ GdnHCl induced denaturation  reported in \cite{tend}\\
$^{b}$ GdnHCl induced denaturation  reported in \cite{CI2}\\
$^{c}$ GdnHCl induced denaturation  reported in \cite{FNIII}\\
$^{d}$ Trifluoroethanol  induced denaturation  reported in \cite{barnase}\\

\newpage

\begin{table}
\caption{Three State Folders }

\begin{tabular}{lccccc}
Protein & pH & C, [Urea] & \(\sigma \) & \(\Omega _{c}({\bf U }
\rightleftharpoons {\bf I }) \) & 
\(\Omega _{c}({\bf I } \rightleftharpoons {\bf N }) \) \\ \hline

Lysozyme$^{a}$  & 2.9  &  --  & 0.21  &  --         &  9.1  \\ \hline
ApoMb$^{b}$     &  --  & 0.0  & 0.41  & 3.4         & 10.5  \\ 
                &  --  & 1.0  & 0.34  & 4.3         & 19.3  \\ 
                &  --  & 1.5  & 0.32  & 3.7         & 24.7  \\ 
                &  --  & 2.0  & 0.32  & 2.7         & 40.2  \\  \hline
ApoMb$^{c}$     & 4.2  &  --  &  --   & 1.2$^{d}$   &  --      \\  
                & 4.2  &  --  &  --   & 5.0$^{e}$   &  --      \\   \hline

\end{tabular}
\end{table}

\noindent
$^{a}$  Urea induced denaturation at pH=2.9 and \(T=20 ^{\circ}C \)
monitored by \(R_{g}\) reported in \cite{lysozyme}\\
$^{b}$  pH induced denaturation at  \(T=0 ^{\circ}C \) 
monitored by CD signal reported in \cite{apoMb}\\
$^{c}$  Urea induced transition at pH=4.2 and \(T=4 ^{\circ}C \)
monitored by fluorescence  reported in \cite{Baldwin97}\\
$^{d}$ 20 mM $Na_{2}SO_{4}$ \\
$^{e}$ 50 mM $NaClO_{4}$

\newpage
\thispagestyle{empty}

\begin{center}

\begin{minipage}{14cm}
\[
\psfig{figure=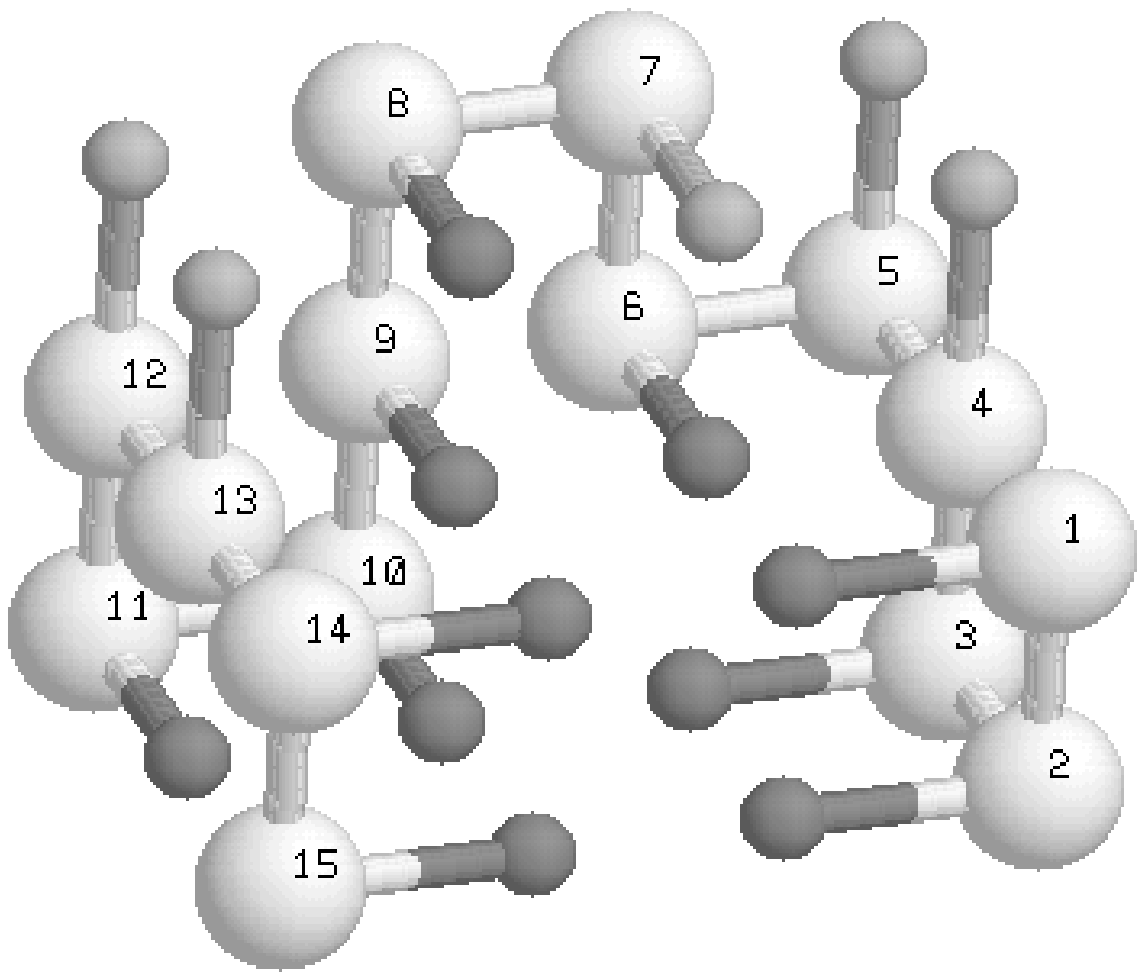,height=10.4cm,width=10.4cm}
\psfig{figure=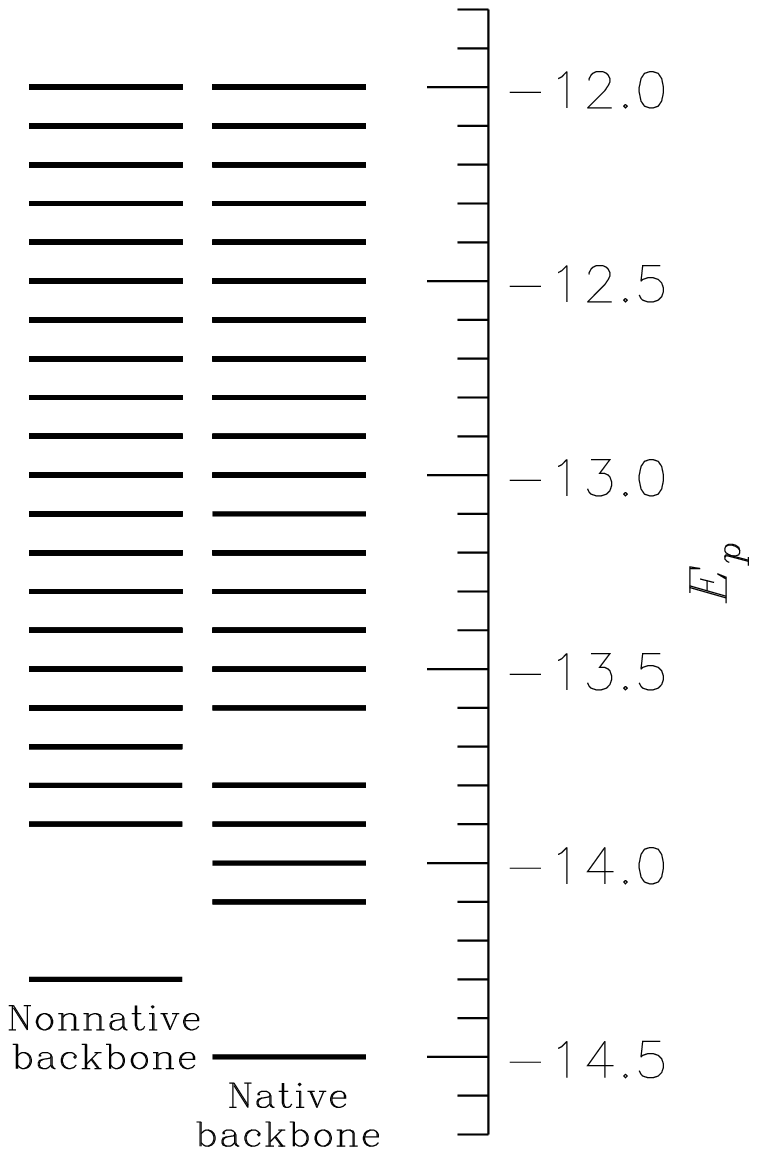,height=9cm,width=10cm}
\]
\end{minipage}

{\Large \bf A: } {\bf WVVEKWHYYVANNAV }

\end{center}
\begin{center}

\enlargethispage{40pt}
\begin{minipage}{14cm}
\[
\psfig{figure=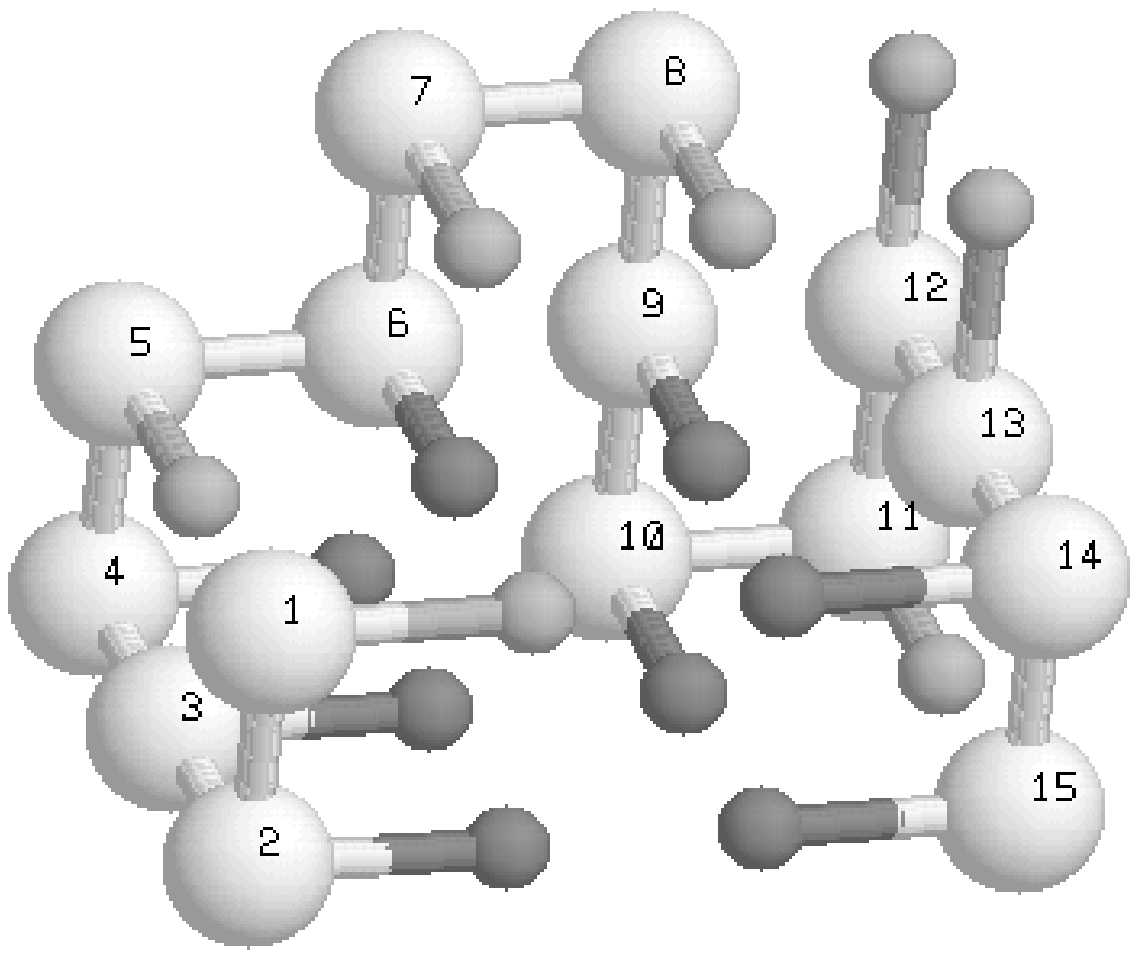,height=10.4cm,width=10.4cm}
\psfig{figure=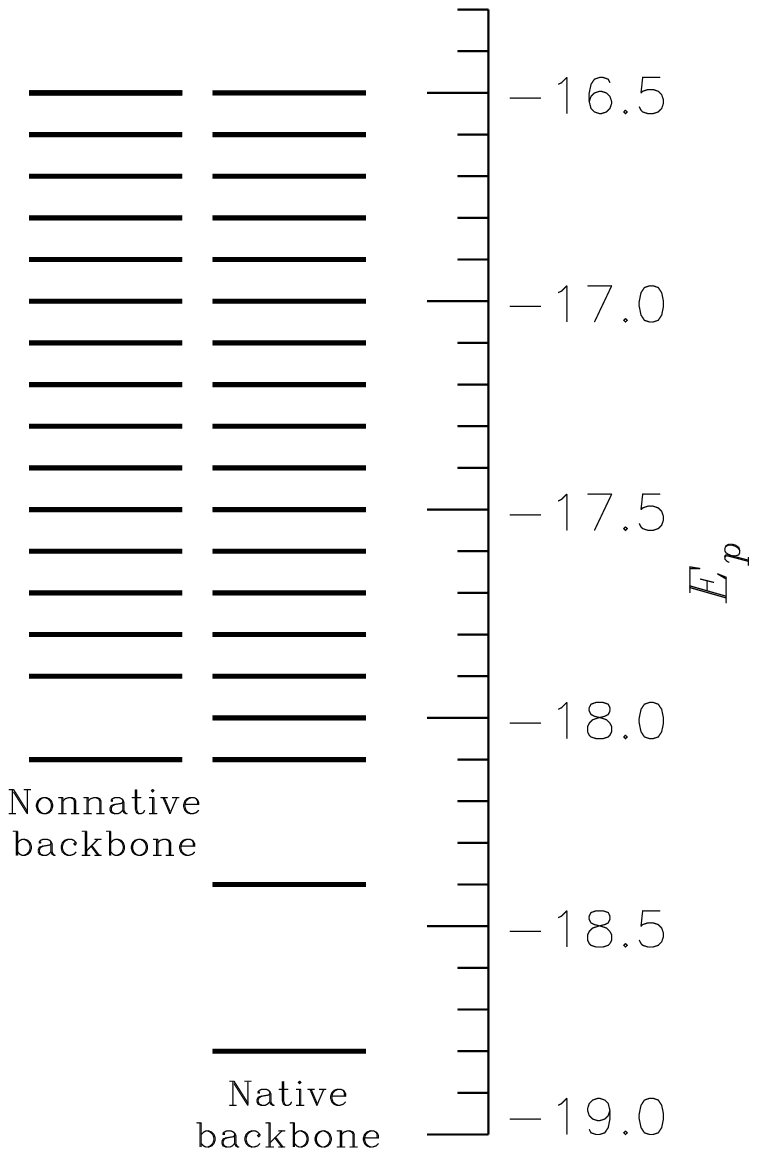,height=9cm,width=10cm}
\]
\end{minipage}

{\Large \bf B: } {\bf 
HCFIGYQRWFRKECM }

{\bf Fig. 1} 
\end{center}

\newpage

\begin{minipage}{5cm}

\end{minipage}

\begin{center}
\begin{minipage}{18cm}
\[
\psfig{figure=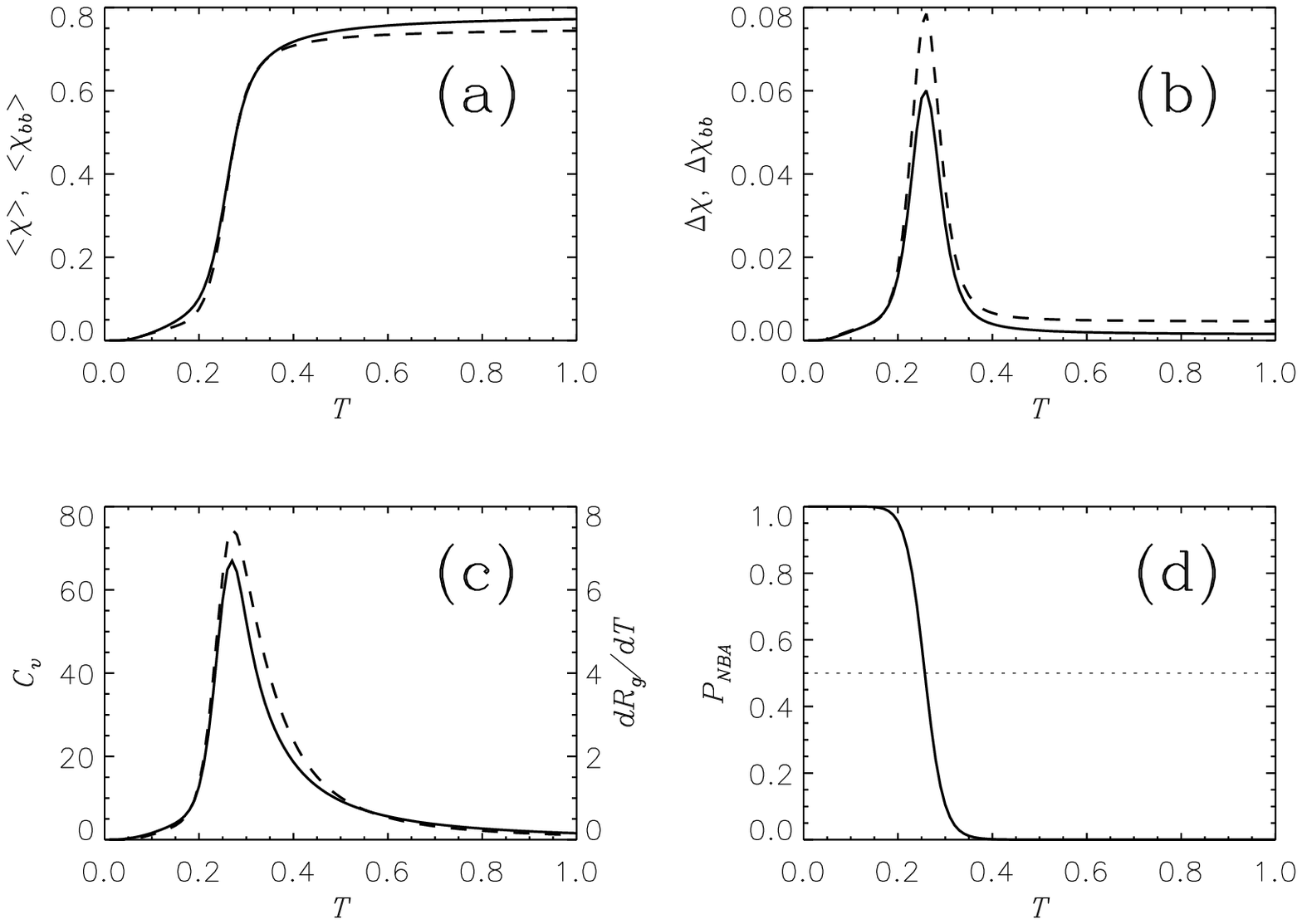,height=17cm,width=19cm}
\]
\end{minipage}

{\bf Fig. 2} 
\end{center}

\newpage

\begin{center}
\begin{minipage}{15cm}
\[
\psfig{figure=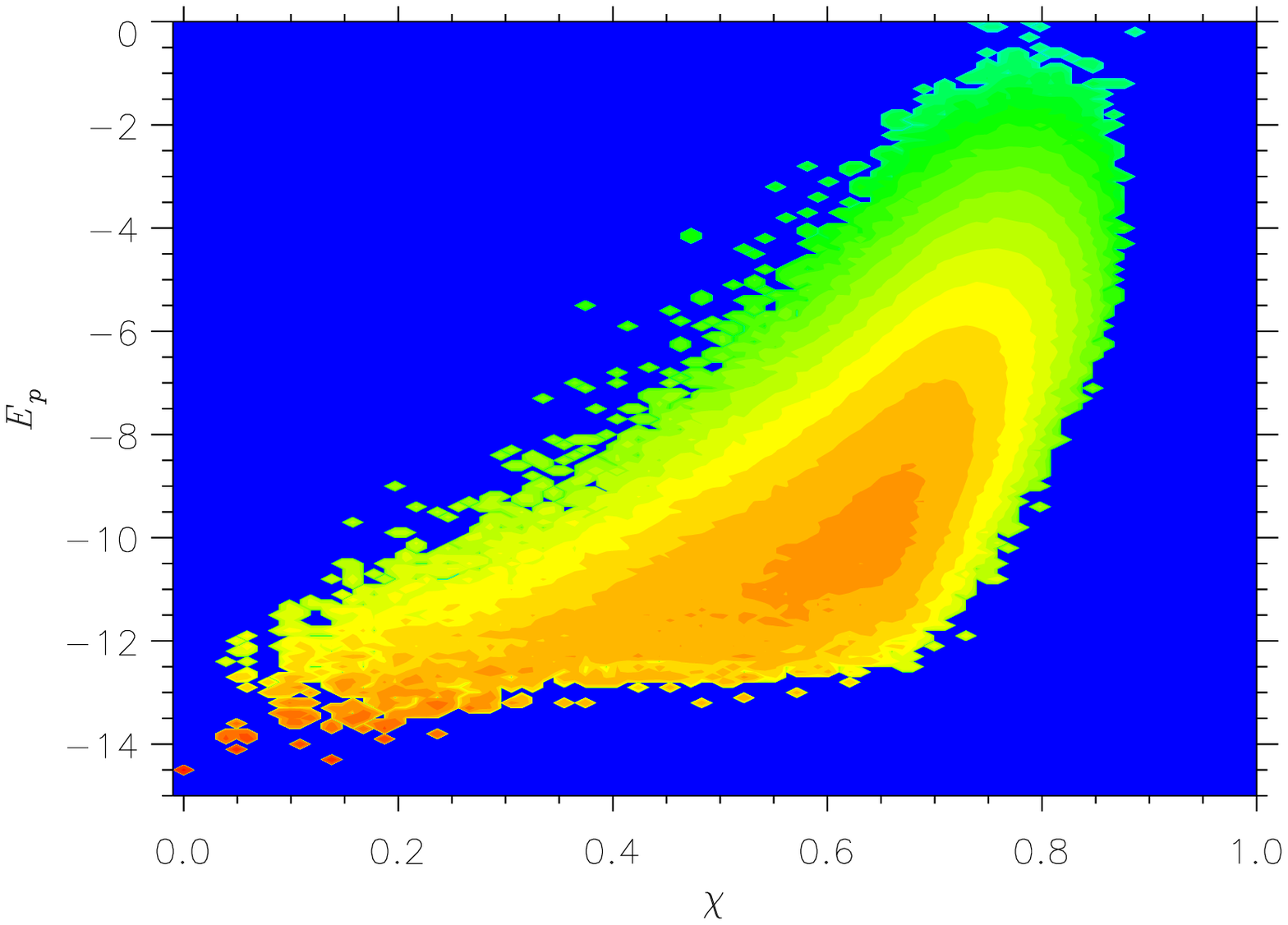,height=9cm,width=13cm}
\]
\end{minipage}
\end{center}

\begin{center}
\begin{minipage}{15cm}
\[
\psfig{figure=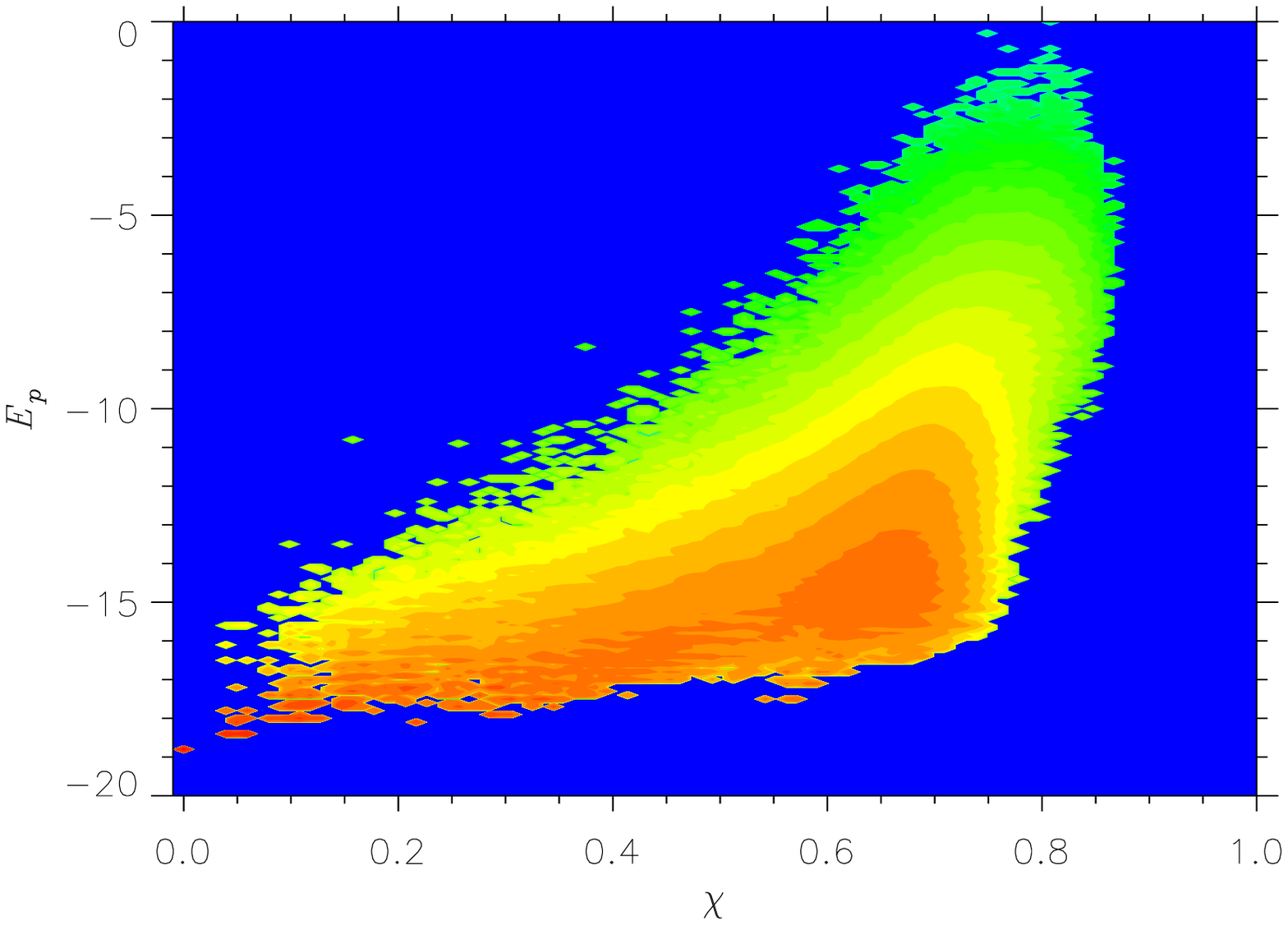,height=9cm,width=13cm}
\]
\end{minipage}

{\bf Fig. 3} 
\end{center}

\newpage 

\begin{center}
\begin{minipage}{15cm}
\[
\psfig{figure=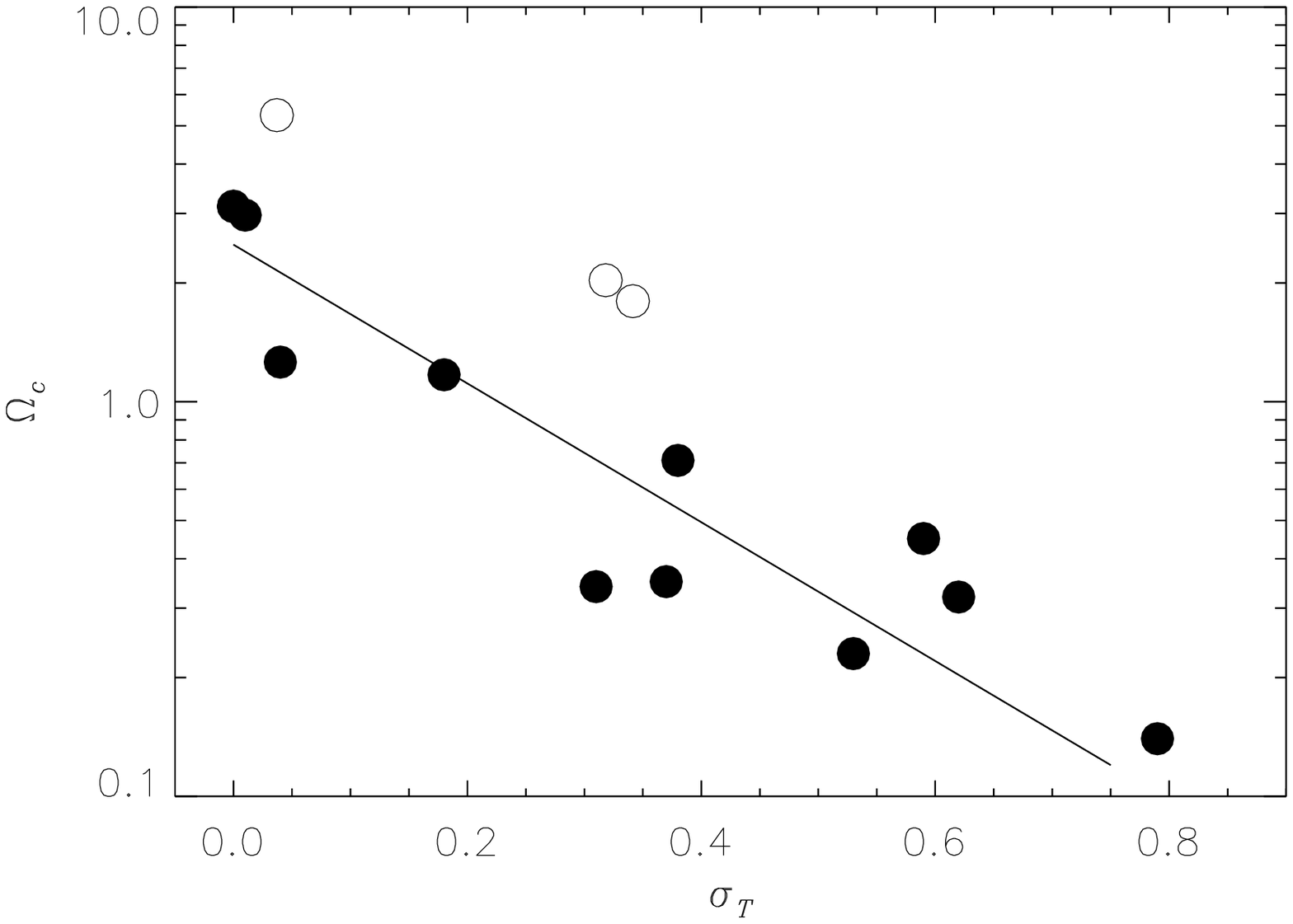,height=9cm,width=12cm}
\]
\end{minipage}

{\bf Fig. 4} 
\end{center}

\begin{center}
\begin{minipage}{15cm}
\[
\psfig{figure=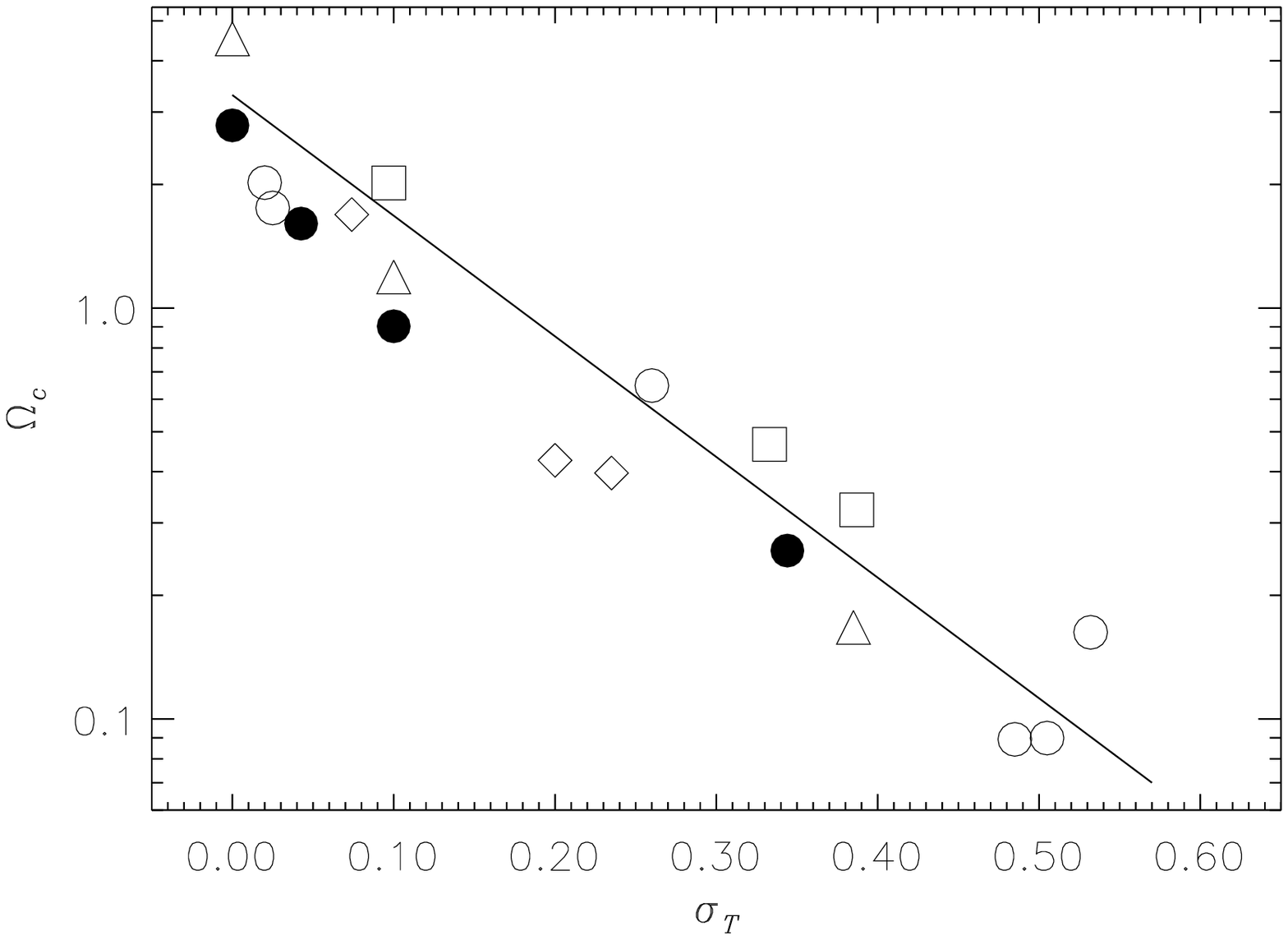,height=9cm,width=12cm}
\]
\end{minipage}

{\bf Fig. 5} 
\end{center}

\newpage 

\begin{center}
\begin{minipage}{15cm}
\[
\psfig{figure=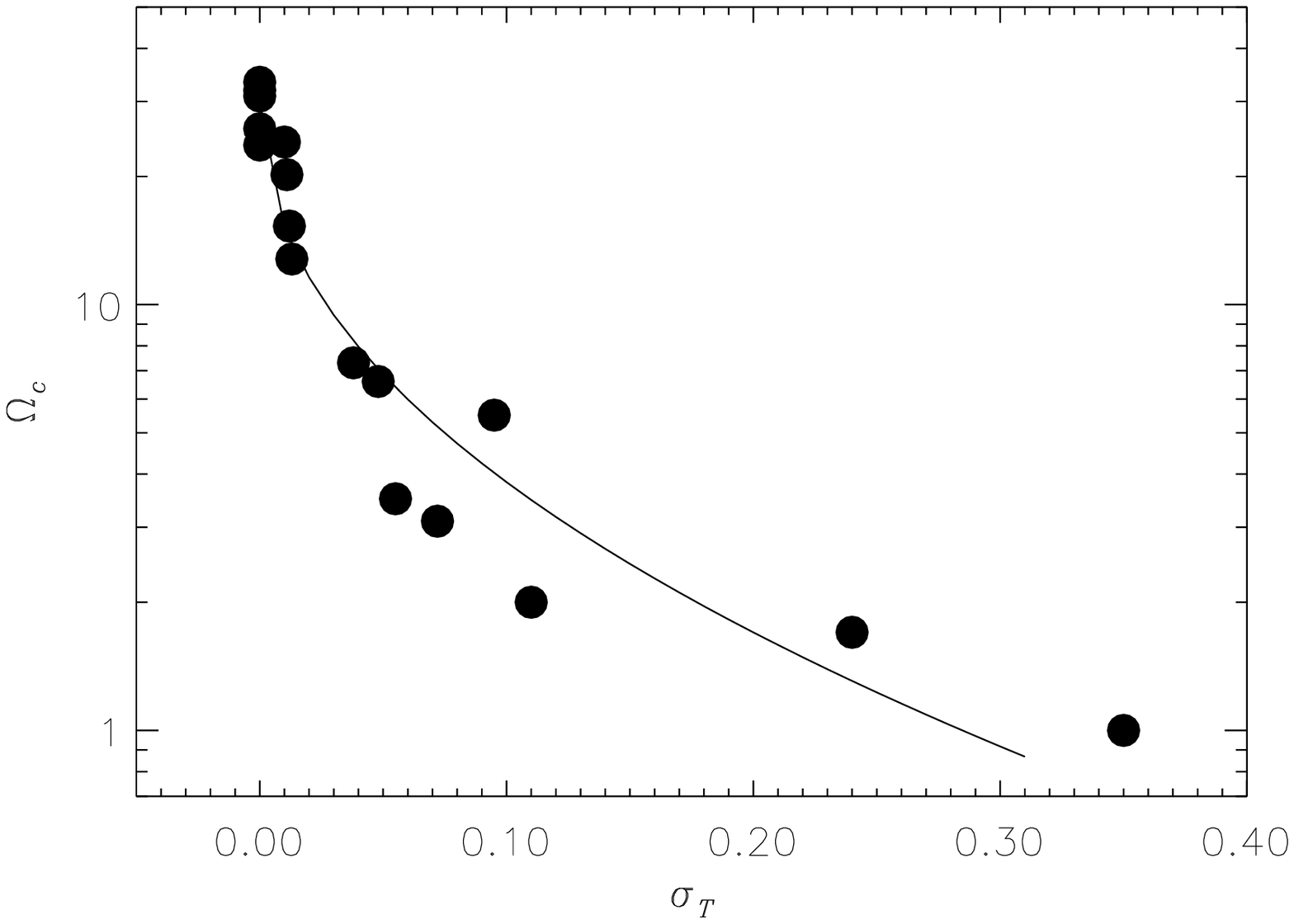,height=9cm,width=12cm}
\]
\end{minipage}

{\bf Fig. 6} 
\end{center}

\begin{center}
\begin{minipage}{15cm}
\[
\psfig{figure=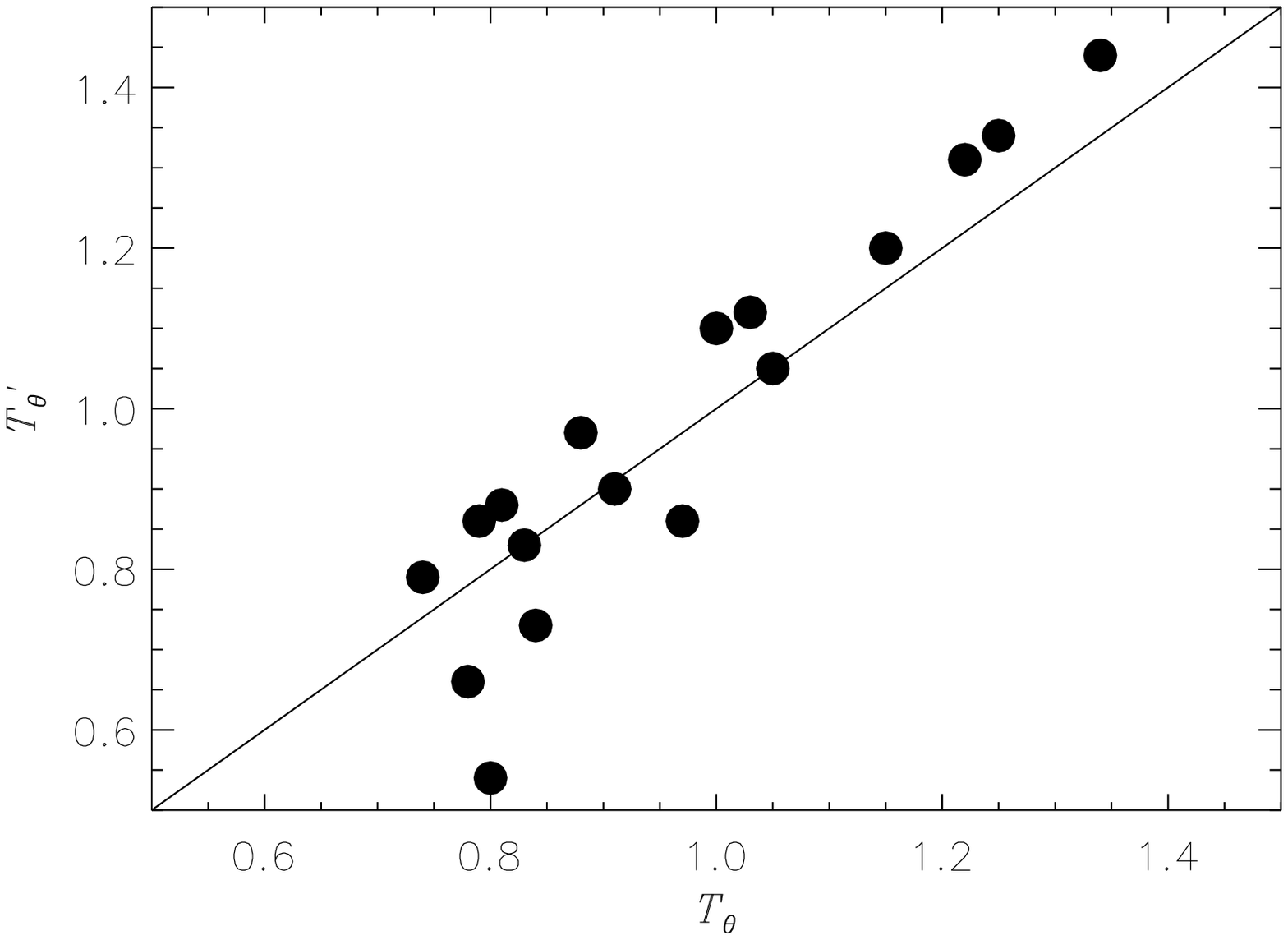,height=9cm,width=12cm}
\]
\end{minipage}

{\bf Fig. 7} 
\end{center}

\newpage 

\begin{center}
\begin{minipage}{15cm}
\[
\psfig{figure=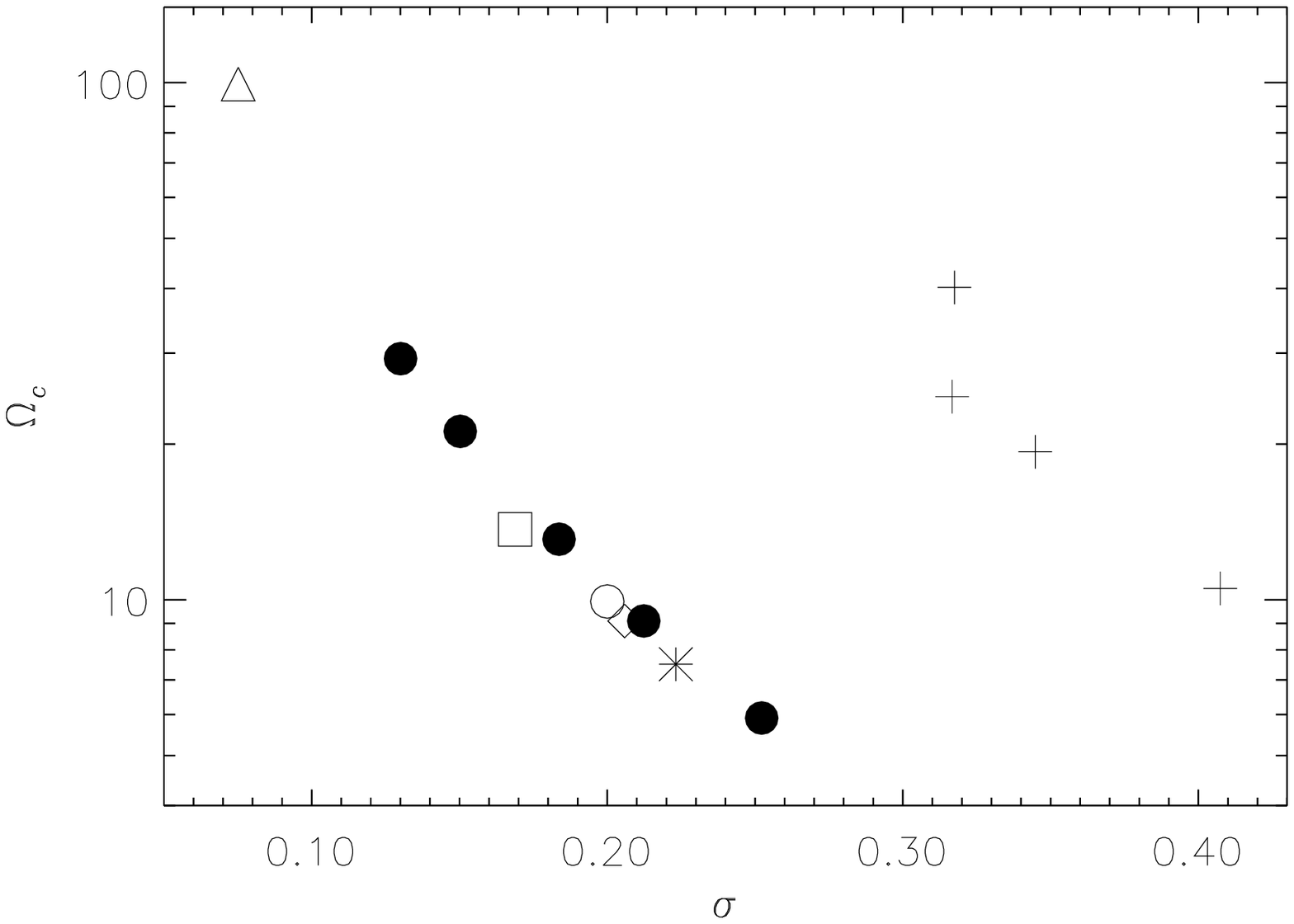,height=9cm,width=12cm}
\]
\end{minipage}

{\bf Fig. 8} 
\end{center}


\begin{references}

\bibitem{Dill95} Dill, K.A., Bromberg, S., Yue, K., Fiebig, K.M., Yee, D.P.,
Thomas, P.D. \& Chan, H.S. (1995). Principles of protein folding
 - A perspective from simple exact models. {\em Protein Sci. } {\bf 4},
561-602. 


\bibitem{Bryn95} Bryngelson, J.D., Onuchic, J.N., Socci, N.D. \& Wolynes,
P.G. (1995). Funnels, pathways and the energy landscape of protein folding: A
synthesis. {\em Proteins Struct. Funct. Genet. } {\bf 21}, 167-195. 

\bibitem{Wol} Wolynes, P.G., Onuchic, J.N. \& Thirumalai, D. (1995). 
Navigating the folding routes. {\em Science } {\bf 267},  1619-1620.


\bibitem{ThirumKlimWood} Thirumalai, D., Klimov, D.K., \& Woodson,
S.A. (1997). Kinetic partitioning mechanisms a unifying theme in the
folding of biomolecules. {\em Theor. Chem. Acct.} {\bf 1}, 23-30. 

\bibitem{DillChan97} Dill, K.A. \&  Chan, H.S. (1997). From Levinthal
to pathways to funnels (1997). {\em Natur. Struct. Biol. } {\bf 4},
10-19. 

\bibitem{Mirny} Mirny, L.A., Abkevich, V. \&  Shakhnovich,
E.I. (1996). Universality and diversity of the protein folding scenarios: a
comprehensive analysis with the aid of a lattice model. {\em Folding \&
Design} {\bf 1}, 103-116. 

\bibitem{Onuchic95} Onuchic, J.N., Wolynes, P.G., Luthey-Schulten,
Z.A. \&  Socci, N.D. (1995). 
Toward an outline of the topography of a realistic protein-folding
funnel. {\em Proc. Natl. Acad. Sci. USA} {\bf 92}, 3626-3630. 

\bibitem{KGS96}  Kolinski, A., Galazka, W., \& Skolnick, J. 
(1996) On the origin of the cooperativity of protein folding:
Implications from model simulations. {\em  Proteins Struct. Funct. 
Genet. } {\bf 26}, 271-287. 

\bibitem{Anfinsen75} Anfinsen, C.B. \& Scheraga, H.A. (1975)
Experimental and theoretical aspects of protein folding. {\em
Adv. Prot. Chem. } {\bf 29}, 205-300. 

\bibitem{Cam93} Camacho, C.J. \&  Thirumalai, D. (1993). Kinetics
and thermodynamics of folding in model proteins. {\em Proc. Natl. Acad. Sci.
USA }{\bf 90}, 6369-6372.


\bibitem{Fukugita93} Fukugita, M., Lancaster, D. \&  Mitchard,
M.G. (1993) Kinematics and thermodynamics of a folding
heteropolymer. {\em Proc. Natl. Acad. Sci. } {\bf 90}, 6365-6368. 

\bibitem{Brooks97} Guo, Z.,  Brooks III, C.L. \& Boczko, E.M.  (1997) 
Exploring the folding free energy surface of a three-helix bundle
protein. 
{\em Proc. Natl. Acad. Sci. } {\bf 94},  10161-10166. 


\bibitem{Karplus97} Zhou, Y. \& Karplus, M. (1997) Folding
thermodynamics of a model three-helix-bundle protein. {\em
Proc. Natl. Acad. Sci. } {\bf 94}, 14429-14432. 



\bibitem{Hao94a} Hao, M.-H. \& Scheraga, H. A. (1994) Monte Carlo simulations
of a first order transition for protein folding. {\em J.  Phys. Chem. }
{\bf 98}, 4940-4948. 
 
\bibitem{Hao94b} Hao, M.-H. \& Scheraga, H. A. (1994) Statistical
thermodynamics of protein folding: Sequence dependence. {\em J.  Phys. Chem. }
{\bf 98}, 9882-9893. 

\bibitem{Richards77} Richards, F. M. (1977) Areas, volumes, packing 
and protein structure. {\em Ann. Rev. Biophys. Bioeng.} {\bf 6},
151-176. 

\bibitem{Richards93} Richards, F. M. \& Lim, W. (1993) An analysis of
packing in the protein folding problem. {\em Q. Rev. Biophys. } {\bf
26}, 423-498. 

\bibitem{BrDill} Bromberg, S. \& Dill, K.A. (1994) Side-chain entropy
and packing in proteins. {\em Protein Sci. } {\bf 3}, 997-1009. 

\bibitem{KGS} Kolinski, A., Godzik, A., \& Skolnick, J. (1993) A
general method for the prediction of the three dimensional structure
and folding pathway of globular proteins: Application to designed
helical proteins. {\em J. Chem. Phys.} {\bf 98}, 7420-7433. 

\bibitem{MJ85}  Miyazawa, S.  \& Jernigan, R.L. Estimation of
effective inter-residue contact energies from protein crystal
structures: quasi-chemical approximation. (1985)  {\em
Macromolecules} {\bf 18}, 534-552. 

\bibitem{MJ96} Miyazawa, S.  \& Jernigan, R.L. Residue-residue
potentials  with a favorable contact pair term and an unfavorable high
packing density term, for simulation and threading. (1996) {\em
J. Mol. Biol. } {\bf 256}, 623-644. 


\bibitem{KlimThirum96} Klimov, D.K. \&  Thirumalai, D. (1996). 
Factors governing the 
foldability of proteins.  {\em Proteins Struct. Funct. Genet. } 
{\bf 26}, 411-441. 


\bibitem{Eisenberg} Bowie, J.U., Luthy, R., \& Eisenberg,
D. (1991). A
method to identify protein sequences that fold into a known
three-dimensional structure. {\em Science } {\bf 253}, 164-170. 

\bibitem{Shakh93} Shakhnovich, E. \&  Gutin, A.M. (1993). 
A new approach to the
design of stable proteins. {\em Protein Eng. }{\bf 6}, 793-800. 

\bibitem{Shakh94} Shakhnovich, E. (1994). 
Proteins with selected sequences fold
into unique native conformation. {\em Phys. Rev. Lett. }{\bf 72}, 3907-3910. 

\bibitem{Veit} Veitshans, T., Klimov, D.K., and Thirumalai, D. (1996).
Protein folding kinetics: Time scales, pathways, and energy
landscapes in terms of sequence dependent properties. {\em
Folding \& Design } {\bf 2}, 1-22. 

\bibitem{Ferrenberg} Ferrenberg, A.M. \& Swendsen,
R. H. (1989). Optimized Monte Carlo data analysis. {\em
Phys. Rev. Lett.} {\bf 63 }, 1195-1198. 

\bibitem{KlimThirumPRL} 
Klimov, D.K. \&  Thirumalai, D. (1996). A criterion that
determines  the foldability of proteins. {\em Phys. Rev. Lett. } 
{\bf 76}, 4070-4073. 



\bibitem{CI2} Jackson, S.E.  \& Fersht, A.R. (1991). Folding of
chymotrypsin inhibitor 2: 1. Evidence for a two-state transition. {\em
Biochem. } {\bf 30}, 10428-10435. 

\bibitem{RNase} Arnold, U. \& Ulbrich-Hofmann, R. (1997). Kinetic and
thermodynamic thermal stabilities of ribonuclease A and ribonuclease
B. {\em Biochem. } {\bf 36}, 2166-2172. 

\bibitem{tend} Schonbrunner, N., Koller, K.-P. \& Kiefhaber, T. (1997)
Folding of the disulfide-bonded \(\beta \)-sheet protein tendamistat:
Rapid two-state folding without hydrophobic collapse. {\em
J. Mol. Biol.} {\bf 268},  526-538. 


\bibitem{FNIII} Plaxco, K. W., Spitzfaden, C., Campbell, I. D. \&
Dobson, C. M. (1997) A comparison of the folding kinetics and
thermodynamics of two homologous fibronectin type III modules.   
{\em J. Mol. Biol.} {\bf 270}, 763-770. 

\bibitem{barnase} Sancho, J., Neira, J. L. \& Fersht, A. R. (1992) A
N-terminal fragment of barnase has residual helical structure similar
to that in a refolding intermediate. 
{\em J. Mol. Biol.} {\bf 224}, 749-758. 


\bibitem{apoMb} Barrick, D. \&  Baldwin, R.L. (1993) Three-state analysis of
sperm whale apomyoglobin folding. {\em Biochem. } {\bf 32},
3790-3796. 

\bibitem{lysozyme} Chen, L., Hodgson, K.O. \& Doniach, S. (1996). A
lysozyme folding intermediate revealed by solution X-ray
scattering. {\em J. Mol. Biol. } {\bf 261}, 658-671. 

\bibitem{Sosnick92} Sosnick, T.R. \& Trewhella, J. (1992) Denaturated
states of ribonuclease A have compact dimensions and residual
secondary structure. {\em Biochem.} {\bf 31}, 8329-8335. 



\bibitem{Baldwin97} Luo, Y., Kay M.S. \& Baldwin, R.L. (1997)
Cooperativity of folding of the apomyoglobin pH 4 intermediate studied
by glycine and proline mutations. {\em Nature Struct. Biol.} {\bf 4},
925-930. 


\bibitem{Balbach} Balbach, J., Forge, V, van Nuland, N. A., Winder,
S. L., Hore, P. J. \& Dobson, C. M. (1995) Following protein folding 
in real-time using NMR-spectroscopy. {\em Nature Struct. Biol. } {\bf
2}, 866-870. 


\bibitem{Mohanty} Mohanty, D., Elber, R., Thirumalai, D., Beglov, D. \&
Roux, B.  (1997) 
Kinetics of Peptide Folding: Computer Simulations of SYPFDV and 
Peptide Variants in Water {\em J. Mol. Biol. } {\bf 272 }, 423-442. 

\bibitem{Finkel} Shakhnovich, E. I. \& Finkelstein, A.V. (1989) Theory
of cooperative transitions in protein molecules. I. Why denaturation
of globular protein is a first order phase transition. {\em
Biopolym. } {\bf 28}, 1667-1680. 


\end{references}
\end{document}